\documentclass[aps,prd,showpacs,showkeys,floatfix,nofootinbib,
twocolumn,10pt]{revtex4-1}
\usepackage[normalem]{ulem}

\usepackage{xcolor}
\usepackage{graphicx}
\usepackage{bm}
\usepackage{amsthm}
\usepackage{amsfonts}
 \usepackage{amsmath,amssymb}
\usepackage{lipsum}
 \usepackage{hyperref}
\DeclareMathOperator{\Tr}{Tr}

\newcommand{\diag}{{\mathrm{diag}}}

\begin{document}
\title{Effective QCD model with consistent quasi-gluon treatment : formulation and application}

\author{Chowdhury Aminul Islam}
\email{caislam.phys@aliah.ac.in}
\affiliation{Department of Physics, Aliah University, II-A/27, Action Area II, Newtown, Kolkata-700160, India}
\author{Munshi G. Mustafa}
\email{Raja Ramanna Chair;\,\, munshigolam.mustafa@saha.ac.in}
\affiliation{Department of Physics, Murshidabad University, 11, Police Reserve Road, Berhampore 742101, India}
\author{Rajarshi Ray}
\email{rajarshi@jcbose.ac.in}
 \affiliation{Department of Physics, Bose Institute, 93/1, A. P. C Road,
 Kolkata - 700009, India}
 \author{Pracheta Singha}
\email{pracheta.singha@gmail.com}
\affiliation{Department of Physics, West University of Timișoara, Bd.~Vasile Pârvan 4, Timișoara 300223, Romania}

%=======================================================================
\begin{abstract}
The Polyakov loop enhanced Nambu-Jona-Lasinio model is reformulated in terms of the gluon quasi-particles in addition to the already existing quark quasi-particles. The formulation goes beyond the saddle point approximation for the gluon sector. The framework provides a physically consistent quasiparticle model for QCD thermodynamics. The ensuing advantages of this formulation is discussed using transport coefficients in the light quark sector.
\end{abstract}

%======================================================================

\maketitle

%----------------------------------------------------------------------
\section{\label{sec:Intro} Introduction}
%----------------------------------------------------------------------
Understanding the phase diagram of quantum chromodynamics (QCD) at finite temperature ($T$) and baryon chemical potential ($\mu_B$) has long  been a central topic of theoretical investigations. 
Experimental observations of spin polarized $\Lambda,\bar{\Lambda}$ hyperons in non central heavy ion collisions~\cite{STAR:2017ckg} have instigated various new directions in these analyses. Finite spin density effects and   non trivial backgrounds, like magnetic filed ($eB$) and rotation ($\Omega$) give  QCD  a rich and complex phase structure  and its theoretical description is dominantly non-perturbative.\\
The  first principle lattice QCD simulation is the most reliable tool in this regard. It has been thoroughly applied to study the phase transition properties at and around the temperature axis of the QCD phase diagram for small $\mu_B$~\cite{Boyd:1996bx, Engels:1999tz,Fodor:2001au,Allton:2002zi,Aoki:2006we,Bazavov:2011nk,Meisinger:1995ih,Levai:1997yx,Karsch:1989pn,CP-PACS:2001hxw,Borsanyi:2012ve,Petreczky:2004xs} , $\Omega$ \cite{Braguta:2022str,Braguta:2023aio},  or spin densities~\cite{Braguta:2025ddq}. However, due to the sign problem a large sector of the QCD phase space is inaccessible  to Lattice simulations. The low energy effective models are particularly insightful in this regimes. These models, based on the basic global 
symmetries of QCD, provide a simplistic approach to understand the partonic interactions.\\ 
 The QCD phase structure is mainly governed by two dynamics \textemdash\, chiral symmetry and confinement. Nonzero chiral condensate, $\sigma$, implies the spontaneous breaking of chiral symmetry and works as an order parameter for the chiral transition. On the other hand, thermal expectation value of the traced Polyakov loop $ (\Phi,\bar{\Phi})$ is associated to the free energy of a heavy quark and serves as an order parameter for the confinement deconfinement phase transition in the $SU(3)$ pure gauge theory.
 In terms of $\Phi,\bar{\Phi}$ an effective potential is often constructed~\cite{Meisinger:1995ih, Pisarski:2000eq, Dumitru:2000in, Pisarski:2002ji} following the Ginsburg-Landau approach~\cite{Landau:1965}, to capture the properties of this first order phase transition of the pure gauge medium. This phase transition is associated with the $Z(3)$ center symmetry which is spontaneously broken leading to a nonzero expectation value of the order parameter in the deconfined phase.

In the presence of quark degrees of freedom, the center symmetry is explicitly broken and the phase transition changes to a crossover. To capture these confinement properties, in the Polyakov extended chiral effective models~\cite{FUKUSHIMA2004277, Megias:2004hj, PhysRevD.73.014019, Ghosh:2006qh, Megias:2006bn, Mukherjee:2006hq,
Ghosh:2007wy, Schaefer:2007pw}, the quarks are considered to be coupled to the background Polyakov field while the 
gluonic contribution to the medium thermodynamics is mimicked through the effective Polyakov potential. These 
Polyakov enhanced chiral effective models have been widely used to study the QCD phase structure in the nonperturbative regions~\cite{Kashiwa:2007hw, Ciminale:2007sr, Fukushima:2008wg, Deb:2009ng, Schaefer:2009ui, Bhattacharyya:2010wp, Bhattacharyya:2010jd, Bhattacharyya:2010ef, Bhattacharyya:2011na,Bhattacharyya:2012rp, Bhattacharyya:2012up, Ghosh:2014zra, Ghosh:2014vja, Bhattacharyya:2014uxa, Bhattacharyya:2015kda, Bhattacharyya:2016jsn, Singha:2017ctc, Bhattacharyya:2017gwt, Singha:2017jmq, Bhattacharyya:2019qhm,Fu:2007xc, Costa:2008yh,Buballa:2008ru,Costa:2009ae,Lourenco:2011buy,
Inagaki:2012re,Inagaki:2012re,Friesen:2011wt, Davoudiasl:2007wf, Sakai:2009vb,Morita:2011eu,Tsai:2008je, Braun:2009gm,Megias:2013xaa,Singha:2025bda,Singha:2025zvh,Singha:2024tpo}. However, 
this prescription misses out on the phase space information of the gluons , restricting its applicability
where gluonic distribution becomes important. ‌

In this work, as an alternative, we propose to 
incorporate the gluonic contribution to the 
chiral effective models such as, Nambu\textemdash 
Jona-Lasinio (NJL)~\cite{Nambu:1961tp, 
Nambu:1961fr, Hatsuda:1994pi, Vogl:1991qt, 
Klevansky:1992qe, Buballa:2003qv}, Quark-Meson 
model~\cite{Jungnickel:1995fp, Berges:1998ha, 
Tetradis:2003qa, Schaefer:2006ds, 
Schaefer:2006sr} etc., via a matrix model structure of Polyakov extended 
quasi-gluon description~\cite{Sasaki:2012bi, 
Ruggieri:2012ny, Islam:2012kv, Alba:2014lda,Pisarski:2016ixt}. Past efforts in this direction were hindered 
due to the unphysical behavior of the thermodynamic quantities below the 
deconfinement temperature, 
$T_d$, as reported in Refs.~\cite{Sasaki:2012bi,Islam:2012kv,Islam:2021qwh}. \\
In our previous work with the pure glue medium~\cite{Islam:2021qwh}, we 
pointed out that these physical inconsistencies 
are the artefacts of using the saddle point 
approximation. There, we proposed an 
alternative approach to evaluate the thermal 
averages and extract thermodynamic quantities. We 
obtained physically consistent results throughout 
the complete relevant temperature range for SU(3) 
pure gauge theory. In the current work, we 
attempt to couple that quasi-gluon description with the 
usual 2-flavor NJL model. Our primary objective 
is to investigate the merit of the proposed 
alternative approach in the presence of quarks. 
This construction gives us a physically consistent prescription to integrate the  gluon matrix model to any chiral effective model. This framework can be employed  to explore further complicated phase structure in presence of different external parameters such as,  rotation, acceleration etc. One can study their effects on the gluonic sector explicitly which is beyond the scope of existing Polyakov extended Chiral effective models. 
 Thus we 
 propose this approach as a more suitable alternative to the existing ones.\\
The main advantage of this quasiparticle framework is that one can obtain the thermal quasi-gluon 
distribution with background 
Polyakov loop. This enables us to explicitly study the   the gluonic contribution to various interaction properties.  
As an illustrative 
example, in this work, we have discussed the transport
coefficients of shear ($\eta$) and bulk ($\zeta$) viscosity extracted within 
this model framework and compared it with the outcome from the usual NJL and PNJL model . 
Although qualitative, this example reveals a 
significant contribution from the quasi-gluon 
sector.\\
The paper is organized as follows. In Sec.~\ref{sec:model} we describe the model framework.  We focus on the gluonic sector in Sec.\ref{ssec:gl_pot} with 
two different effective potential description, the polynomial potential approach in Sec.\ref{sssec:pol_appr} and the Quasi-particle approach in Sec.\ref{sssec:qp_appr}.
Next in Sec. \ref{ssec:PNJL}  we integrate the gluonic description to the quark sector.  We discuss the saddle point approximation and the alternative integral approach, proposed in Ref.~\cite{Islam:2021qwh}, to obtain the physical observables.  We devote 
Sec.~\ref{sec:result} to mainly showcase the 
consistency of our model in successfully 
reproducing the thermodynamic observables known 
from the lattice QCD study. Exploring the 
parameter space in Sec.\ref{ssec:param} we present our numerical results for order parameters and thermodynamic observables in Sec.\ref{ssec:OP} and Sec.\ref{ssec:Therm} respectively. 
Sec.\ref{sec:App} deals with the application 
part of our approach. We obtain the distribution function in Sec. \ref{ssec:distri} and discuss the 
transport coefficients in Sec. \ref{ssec:trans}. 
Finally, 
we summarise and conclude in 
Sec.\ref{sec:discussion}.

%-----------------------------------------------------------------------
\section{\label{sec:model} Model details}
In this section, we present the details of the effective model proposed. To put things into perspective, we will briefly present the standard approach of treating the gluonic and Polyakov loop contributions side by side in subsection~\ref{ssec:gl_pot}. We describe the standard polynomial approach~(\ref{sssec:pol_appr}) as well as the gluon quasi-particle approach~(\ref{sssec:qp_appr}), which was discussed in Refs.~\cite{Sasaki:2012bi, Islam:2012kv, Islam:2021qwh}. Although quark quasi-particles can be consistently treated using the chiral effective models like NJL model  ~\cite{Nambu:1961tp, Nambu:1961fr}, the quasi-gluon model developed in Refs.~\cite{Sasaki:2012bi,Islam:2012kv} seemed to be incomplete. In Ref. \cite{Islam:2021qwh} we developed a novel scheme to remove the anomaly of the description of gluon quasiparticles. Using that framework we are now in a position to consider both the quasi-gluons and quasi-quarks to build a quasi-particle picture in the PNJL framework, as described in subsection~\ref{ssec:PNJL}. We shall discuss both the standard PNJL approach with polynomial gluonic potential~(\ref{sssec:std_appr}) and our alternative approach with quasi-gluon potential~(\ref{sssec:alt_appr}).
%-----------------------------------------------------------------------
\subsection{\label{ssec:gl_pot}Effective potential for gluons}
The traced Polyakov loop, or the normalized characters of Polyakov loop are defined as,
%%%%%%%%%%%%%%%%%%%%%%%%%%%%%%%%%%%%%%%%%%%%%%%%%%%%%%%%%
\begin{equation}
\Phi=\frac{1}{N_c}\Tr\hat{L}_F; \hspace{5mm}    
\bar{\Phi}=\frac{1}{N_c}\Tr\hat{L}^\dagger_F~
\label{funcharacter}
\end{equation}
%%%%%%%%%%%%%%%%%%%%%%%%%%%%%%%%%%%%%%%%%%%%%%%%%%%%%%%%%%
and,
%%%%%%%%%%%%%%%%%%%%%%%%%%%%%%%%%%%%%%%%%%%%%%%%%%%%%%%%%%%%%%%%%%%%%%%%
\begin{equation}
\Phi_A=\frac{1}{N_c^2-1}\Tr\hat{L}_A
=\frac{1}{N_c^2-1}\left({N_c^2\Phi\bar{\Phi}-1}\right)~,
\label{adjcharacter}
\end{equation}
%%%%%%%%%%%%%%%%%%%%%%%%%%%%%%%%%%%%%%%%%%%%%%%%%%%%%%%%%%%%%%%%%%%%%%%%
for fundamental ($\hat{L}_F$) and adjoint ($\hat{L}_A$) representation, respectively, with $N_c=3$ for $SU(3)$ Yang-Mills theory. In the Polyakov gauge, for general representation $R$, $\hat{L}_R$ can be written in the diagonal form as,
\begin{equation}
    \hat{L}_R=\mathrm{exp}[i(A_4^3 
    \lambda_R^3+A_4^8\lambda_R^8)/T],
\end{equation}
where $A_4$ is the temporal component of the background Euclidean gauge field and $ \lambda_R^3$, 
$\lambda_R^8$ are the diagonal generators of $SU(3)$. One can also represent these Polyakov matrices in terms of 
two class parameters of $SU(3)$, $\theta_1$ and $\theta_2$, as~\cite{Sasaki:2012bi},
%%%%%%%%%%%%%%%%%%%%%%%%%%%%%%%%%%%%%%%%%%%%%%%%%%%%%%%%%%%%%%%%%%%%%%%%
\begin{equation}
\hat{L}_F = \diag(e^{i\theta_1},e^{i\theta_2},e^{-i(\theta_1+\theta_2)})~,
\label{funmatrix}
\end{equation}
%%%%%%%%%%%%%%%%%%%%%%%%%%%%%%%%%%%%%%%%%%%%%%%%%%%%%%%%%%%%%%%%%%%%%%%
and,
%%%%%%%%%%%%%%%%%%%%%%%%%%%%%%%%%%%%%%%%%%%%%%%%%%%%%%%%%%%%%%%%%%%%%%%%
\begin{multline}
\hat{L}_A = \diag\left(1, 1, e^{i(\theta_1-\theta_2)},
e^{-i(\theta_1-\theta_2)}, e^{i(2\theta_1+\theta_2)}, 
\right.\\ 
\left.e^{-i(2\theta_1+\theta_2)}, 
e^{i(\theta_1+2\theta_2)},
e^{-i(\theta_1+2\theta_2)}\right)~.
\label{eqadjLphi}
\end{multline}
%%%%%%%%%%%%%%%%%%%%%%%%%%%%%%%%%%%%%%%%%%%%%%%%%%%%%%%%%%%%%%%%%%%%%%%%
\noindent
In most of the present literature the effective potential describing the gluon thermodynamics are constructed in terms of these normalized characters and often include a Vandermonde determinant term motivated from the $SU(3)$ Haar measure. For general $SU(N_c)$, the  Haar measure $d\mu$ can be written in terms of the distribution of the eigenvalues as,
%%%%%%%%%%%%%%%%%%%%%%%%%%%%%%%%%%%%%%%%%%%%%%%%%%%%%%%%%%%%%%%%%%%%%%%%%%%%%
\begin{align}
    \int d\mu&=\frac{1}{N_c!}\left(\prod_{i=1}^{N_c} 
    \int_0^{2\pi}\frac{d\theta_i}{2\pi}\right)
    \delta\left(\sum_i \theta_i\right)\prod_{i<j} |
    e^{i\theta_i}-e^{i\theta_j}|^2\nonumber\\&=1~.
\end{align}
%%%%%%%%%%%%%%%%%%%%%%%%%%%%%%%%%%%%%%%%%%%%%%%%%%%%%%%%%%%%%%%%%%%%%%%%%%%%%%
For $SU(3)$ it simplifies to,
%%%%%%%%%%%%%%%%%%%%%%%%%%%%%%%%%%%%%%%%%%%%%%%%%%%%%%%%%%%%%%%%%%%%%%%%%%%%%%%%
\begin{align}
 \frac{1}{3!} \int_0^{2\pi}\int_0^{2\pi}\frac{d\theta_1}
 {2\pi}\frac{d\theta_2}{2\pi}\mathrm{Det}_{\mathrm{VDM}}
 [\theta_1,\theta_2]=1~,
\end{align}
%%%%%%%%%%%%%%%%%%%%%%%%%%%%%%%%%%%%%%%%%%%%%%%%%%%%%%%%%%%%%%%%%%%%%%%%%%%%%%%%%%
where, $\mathrm{Det}_{\mathrm{VDM}}$ is the Vandermonde determinant term, given as,
%%%%%%%%%%%%%%%%%%%%%%%%%%%%%%%%%%%%%%%%%%%%%%%%%%%%%%%%%
\begin{align}
   \mathrm{Det}_{\mathrm{VDM}}&=64 \sin^2\frac{(\theta_1-
   \theta_2)}{2}\sin^2\frac{(2\theta_1+\theta_2)}{2}
   \sin^2\frac{(\theta_1+2\theta_2)}{2}\nonumber \\&=
   27[1-6\bar{\Phi}\Phi+4(\bar{\Phi}^3+
   \Phi^3)-3(\bar{\Phi}\Phi)^2]~.
\end{align}
%%%%%%%%%%%%%%%%%%%%%%%%%%%%%%%%%%%%%%%%%%%%%%%%%%%%%%%%%%%%%%%%%%%%%%%%%%%%%%%%%%

Including this Vandermonde determinant term the effective potentials can be written as,
%%%%%%%%%%%%%%%%%%%%%%%%%%%%%%%%%%%%%%%%%%%%%%%%%%%%%%%%%%%%%%%%%%%%%%%%%%%%%%%%%%
\begin{align}
   \Omega_{\mathrm{Poly}}&=U(\Phi,\bar{\Phi})+
\kappa T^4\ln{\mathrm{Det}_{\mathrm{VDM}}}
\label{eqpoly}\\
\Omega_{\mathrm{qp}}&=\Omega_{\mathrm{gqp}}+\kappa 
T^4\ln{\mathrm{Det}_{\mathrm{VDM}}}~,
\label{eqquasi}
\end{align}
%%%%%%%%%%%%%%%%%%%%%%%%%%%%%%%%%%%%%%%%%%%%%%%%%%%%%%%%%%%%%%%%%%%%%%%%%%%%%%%%%
where, $\kappa$ is a phenomenological parameter and $U(\Phi,\bar{\Phi})$ and $\Omega_{\mathrm{gqp}}$ are, respectively, the Landau polynomial and quasi-particle potentials for gluonic degrees of freedom. 
%**************************************************************************************
\subsubsection{\label{sssec:pol_appr}Polynomial potential approach}
%*********************************************************************
In this approach, a $Z(3)$ symmetric polynomial potential is constructed in terms of $\Phi,\bar{\Phi} $ and the first order phase transition is captured through the temperature dependent parameters, $\alpha_i$, following the usual 
Ginsburg-Landau approach~\cite{Landau:1965}.  $U(\Phi,\bar{\Phi})$ can be written as~\cite{FUKUSHIMA2004277, Megias:2004hj, PhysRevD.73.014019, Ghosh:2006qh, Megias:2006bn, Mukherjee:2006hq, Ghosh:2007wy, Schaefer:2007pw},
%%%%%%%%%%%%%%%%%%%%%%%%%%%%%%%%%%%%%%%%%%%%%%%%%%%%%%%%%%%%%%%%%%%%%%%%
\begin{equation}
\frac{U(\Phi,\bar{\Phi})}{T^4}=-\frac{\alpha_1(T)}{2}\Phi 
\bar{\Phi}
-\frac{\alpha_2}{6}(\Phi^3+\bar{\Phi}^3)+\frac{\alpha_3}
{4}(\Phi 
\bar{\Phi})^2~,
\label{equpoly}
\end{equation}
%%%%%%%%%%%%%%%%%%%%%%%%%%%%%%%%%%%%%%%%%%%%%%%%%%%%%%%%%%%%%%%%%%%%%%%%
where 
\begin{equation}
    \alpha_1(T)=b_0+b_1\left(\frac{T_d}{T}\right)
    +b_2\left(\frac{T_d}{T}\right)^2+b_3\left(\frac{T_d}
    {T}\right)^3~.
    \label{eqalpha1}
\end{equation}
%%%%%%%%%%%%%%%%%%%%%%%%%%%%%%%%%%%%%%%%%%%%%%%%%%%%%%%%%%%%%%%%%%%%%%%%%
In Eq.\eqref{equpoly} and Eq.\eqref{eqalpha1}, $b_0,b_1,b_2,b_3,\alpha_2,\alpha_3$ and $T_d$ are the model parameters 
that are fixed to reproduce the Yang-Mills thermodynamics evaluated from the lattice pure gauge simulation~\cite{PhysRevD.73.014019}. 
In our numerical analysis to reproduce results for usual PNJL model, we have used $\Omega_\mathrm{Poly}$ with the parameters given 
in table.\ref{table_u}, taken from Ref.~\cite{Ghosh:2007wy,PhysRevD.73.014019}.
%%%%%%%%%%%%%%%%%%%%%%%%%%%%%%%%%%%%%%%%%%%%%%%%%%%%%%%%%%%%%%%%%%%%%%%%
\begin{table}
	\centering
	\begin{tabular}{|c|c|c|c|c|c|c|c|}
		\hline
$b_0$&$b_1$&$b_2$&$b_3$&$\alpha_2$&$\alpha_3$ & 
		$T_d$ (GeV)& $\kappa$\\[1 ex]
		%		\hline
		\hline
		6.75 & -1.95 & 2.625 & -7.44 & 0.75 & 7.5 & 0.270& 
		0.1\\[1 ex]
		\hline
	\end{tabular}
	\caption{Parameters for the gluonic potential in the polynomial form.}
	\label{table_u}
\end{table}

%%%%%%%%%%%%%%%%%%%%%%%%%%%%%%%%%%%%%%%%%%%%%%%%%%%%%%%%%%%%%%%%%%%%%%%%
Though this approach has drawn incredible attention and given us a wealth of information over the last two decades, it has a serious shortcoming in that it cannot estimate any dynamical properties of the gluon sector. This would be only possible if a quasi-particle approach for gluons can be developed.

%**********************************************************************
\subsubsection{\label{sssec:qp_appr}Quasi-particle approach}
%*********************************************************************
In the quasi-particle model~\cite{Ruggieri:2012ny,Alba:2014lda,Islam:2012kv,Sasaki:2012bi,Islam:2021qwh}, the gluon quasi-particles are considered to be swarming in the background of Polyakov field. The thermodynamic potential, $\Omega_{\mathrm{gqp}}$ can be written as,
 %%%%%%%%%%%%%%%%%%%%%%%%%%%%%%%%%%%%%%%%%
\begin{eqnarray}
\Omega_{\mathrm{gqp}}=
&=& {2T\int{\frac{d^3p} {(2\pi)^3} 
\ln\det{\big(1-\hat{L}_A e^{-\frac{|\vec{p}|}{T}}\big)}}} 
\nonumber 
\\
&=& {2T\int{\frac{d^3p} {(2\pi)^3}
\ln\left(1+\sum^{8}_{n=1} a_n e^{-\frac{n |\vec{p}|}
{T}}\right)}}~,
\label{eqsimpleg}
\end{eqnarray}
%%%%%%%%%%%%%%%%%%%%%%%%%%%%%%%%%%%%%%%%%%%%%%%%%%%%%%%%%%%%%%%%%%%%%%%%
where $\hat{L}_A$ is given in Eq.~\eqref{eqadjLphi}. The coefficients $a_n$, for $n= 1 \cdots 8$, are obtained after performing the colour trace and can be written in terms of $\Phi,\bar{\Phi}$ as~\cite{Islam:2012kv,Sasaki:2012bi},
%%%%%%%%%%%%%%%%%%%%%%%%%%%%%%%%%%%%%%%%%%%%%%%%%%%%%%%%%%%%%%%%%%%%%%%%
\begin{eqnarray}
&a_8 &= 1\nonumber;~~a_1 = a_7=1-9\bar{\Phi}\Phi;\nonumber\\
&a_2 &= a_6=1-27\bar{\Phi}\Phi +27(\bar{\Phi}^3+\Phi^3);\nonumber\\
&a_3 &= a_5=-2+27\bar{\Phi}\Phi-81(\bar{\Phi}\Phi)^2;\nonumber\\
&a_4 &= 2[-1+9\bar{\Phi}\Phi-27(\bar{\Phi}^3+\Phi^3)+81(\bar{\Phi}\Phi)^2]~.
\label{eqpolya_distri}
\end{eqnarray}
%%%%%%%%%%%%%%%%%%%%%%%%%%%%%%%%%%%%%%%%%%%%%%%%%%%%%%%%%%%%%%%%%%%%%%%%%%%
\\
Thus, in this quasi-particle model, the thermal distribution function of quasi-gluons is modified through the background Polyakov field. Consequently, the confinement properties can be realized via the statistical suppression as at temperatures below $T_d$, $Z(3)$ symmetry is restored and $\langle \Phi\rangle,\langle\bar{\Phi}\rangle=0$.

However, employing this quasi-particle model to study QCD  properties faced a major drawback, as some thermal observables became physically inconsistent at temperatures below the deconfinement temperature ($T_d$)~\cite{Sasaki:2012bi,Ruggieri:2012ny,Islam:2012kv,Alba:2014lda}. Various remedies were suggested by the authors to fix the issue. For example, in Ref. \cite{Sasaki:2012bi} a hybrid approach was proposed including the glueballs for low temperatures, a bag pressure term was introduced along with the quasi-gluon In Ref.~\cite{Alba:2014lda} and in Ref~\cite{Ruggieri:2012ny}, an additional Polyakov potential term was introduced. Although, such modifications bring in additional features to the model ensuring an overall physical consistency, none directly addresses the issue of quasi-particle thermodynamics being negative at low temperatures.

Following these, in our previous work~\cite{Islam:2021qwh,Singha:2021wpi}, we have identified that the unphysical behaviour is the artefact of using the mean-field approximation employing the saddle point solutions. 
Given the partition function,
%##############################################################################################
\begin{align}
Z_\mathrm{g}
&= \int{\mathcal{D}\theta_1\mathcal{D}\theta_2 
\mathrm{exp}\left[-\frac{1}{T}\int d^3x 
\Omega_{\mathrm{gqp}}[\theta_1({\bf x}),\theta_2({\bf x})]
\right]} \nonumber 
\\
&= \int\prod_{\bf x} \frac{1}{24\pi^2}
d\theta_1({\bf x}) d\theta_2({\bf x}) \mathrm{Det}
_{\mathrm{VDM}}
\nonumber \\
&~~~~~~~~~\mathrm{exp}\left[-\frac{1}{T}\int d^3x 
\Omega_{\mathrm{gqp}}[\theta_1({\bf x}),\theta_2({\bf x})]
\right],
\label{eqzpart1}
\end{align}
%##############################################################################################
where $\Omega_{\mathrm{gqp}}$ is given by Eq.~\eqref{eqsimpleg}, we noted that the Polyakov loop and consequently, the thermodynamic potential are oscillatory functions of $\theta_1$ and $\theta_2$. We inferred that the configurations away from the saddle point solution might have a significant contribution. So, instead of the saddle point approximation, we introduced a slightly different approach. In the saddle point approximation only the most important constant field configuration is picked up. Instead we allowed for all possible constant field configurations to contribute. This was achieved through an integration of the partition function for all constant values of $\theta_1$ and $\theta_2$ for the finite periodic interval 0 to $2\pi$, with the corresponding weight factors given by the thermodynamic potential. The integration was plausible as the effective action did not contain any derivative term in two class parameters and the configuration space could be split up into $\mathrm{N} \rightarrow \infty$ equivalent and independent points. The partition function is then simplified to,
%%%%%%%%%%%%%%%%%%%%%%%%%%%%%%%%%%%%%%%%%%%%%%%%%%%%%%%%%%%%%%%%%%%%%%%%
\begin{eqnarray}
Z_\mathrm{g} &=& z_\mathrm{g}^N
\label{part_coarse}
\end{eqnarray}
%%%%%%%%%%%%%%%%%%%%%%%%%%%%%%%%%%%%%%%%%%%%%%%%%%%%%%%%%%%%%%%%%%%%%%%%
where
%%%%%%%%%%%%%%%%%%%%%%%%%%%%%%%%%%%%%%%%%%%%%%%%%%%%%%%%%%%%%%%%%%%%%%%% 
\begin{eqnarray}
z_\mathrm{g} = \int \frac{1}{24\pi^2}d\theta_1 d\theta_2
\mathrm{Det}_{\mathrm{VDM}} \mathrm{exp}\left[-\frac{v}{T}
\omega_{gqp}[\theta_1,\theta_2]\right],
\label{eq.z}
\end{eqnarray}
%%%%%%%%%%%%%%%%%%%%%%%%%%%%%%%%%%%%%%%%%%%%%%%%%%%%%%%%%%%%%%%%%%%%%%%%%%%%%
with, $v$ as a model parameter with the dimension of volume. 

The approach is short of a full field theoretic path integration (like lattice QFT) in that the spatial derivatives were neglected. So this is a model framework having various phenomenological parameters, but shows a way forward where mean field analysis is physically untenable.

Once the partition function is obtained, the thermodynamic quantities were extracted following the usual thermodynamic relations. We obtained physically consistent results throughout the complete temperature range and introducing a thermal quasi-gluon mass, a good quantitative agreement to the lattice $SU(3)$ pure gauge results were 
achieved~\cite{Islam:2021qwh,Singha:2021wpi}.
%****************************************************************************
\subsection{\label{ssec:PNJL} Coupling with the chiral model : PNJL model}
%****************************************************************************
We are now in a position to develop a PNJL framework where both the quarks and gluons have their respective thermodynamics given in terms of the corresponding quasi-particles, with the interactions being incorporated through their coupling to the quark and gluon condensates. We first briefly mention the standard Polyakov loop potential and mean field approach in subsection \ref{sssec:std_appr} and subsequently present our main formalism in subsection \ref{sssec:alt_appr}.

\subsubsection{\label{sssec:std_appr} The standard approach}

Nambu\textemdash Jona-Lasinio model~\cite{Nambu:1961tp, Nambu:1961fr, Hatsuda:1994pi, Vogl:1991qt, Klevansky:1992qe, Buballa:2003qv}, is a chiral effective model where the gauge field is integrated out and replaced by the local four-point interaction of quark degrees of freedom. To incorporate the confinement properties in this chiral model, the quarks are considered to be coupled to the static background gauge field and gluon thermodynamics is captured in terms of the effective Polyakov potential like the polynomial potential \eqref{equpoly} with or without the Vandermonde determinant term given in Eq. \eqref{eqpoly}. In the following, we denote any such generic Polyakov-loop effective potential by $U(\Phi,\bar{\Phi})$. The Lagrangian of the two flavour ($\psi=(u,d)$) Polyakov-Nambu\textemdash Jona-Lasinio model is given as,
%%%%%%%%%%%%%%%%%%%%%%%%%%%%%%%%%%%%%%%%%%%%%%%%%%%%%%%%%%%%%%%%%%%%%%%%
\begin{equation}
\mathcal{L}=\bar{\psi}(i\gamma_\mu D^\mu-\hat{m}_0) \psi + 
G[(\bar{\psi}\psi)^2+(\bar{\psi}i\gamma_5 \tau \psi)^2]-
U(\Phi,\bar{\Phi})~,
\label{pnjl1}
\end{equation}
%%%%%%%%%%%%%%%%%%%%%%%%%%%%%%%%%%%%%%%%%%%%%%%%%%%%%%%%%%%%%%%%%%%%%%%%
where, $\hat{m}_0$ is the two flavour current quark mass matrix; with the isospin symmetry, $\hat{m}_0=\mathrm{Diag}
[m_u,m_d]=\mathrm{Diag}[m_0,m_0]$. The covariant derivative is given by,
%%%%%%%%%%%%%%%%%%%%%%%%%%%%%%%%%%%%%%%%%%%%%%%%%%%%%%%%%%%%%%%%%%%%%%%%
\begin{equation}
D^\mu=\partial^\mu-i A^\mu \text{ and } 
A^\mu=\delta_{\mu0}A^0~.
\label{pnjl2}
\end{equation}
%%%%%%%%%%%%%%%%%%%%%%%%%%%%%%%%%%%%%%%%%%%%%%%%%%%%%%%%%%%%%%%%%%%%%%%%
 
The thermodynamic potential obtained from Eq.~\eqref{pnjl1} can be written as \cite{Pisarski:2000eq,Dumitru:2000in,FUKUSHIMA2004277,Megias:2004hj,Osipov:2005tq, Osipov:2005sp,Megias:2006bn, 
PhysRevD.73.014019,Osipov:2006ns, Osipov:2006ns, Kashiwa:2006rc,Mukherjee:2006hq,Ghosh:2006qh,Kashiwa:2007hw, Ghosh:2007wy, Ciminale:2007sr,Fukushima:2008wg, Hiller:2008nu,Deb:2009ng,Bhattacharyya:2010ef,Bhattacharyya:2010jd, Bhattacharyya:2010wp, Bhattacharyya:2011na, Deb:2011en, Bhattacharyya:2012up,Bhattacharyya:2012rp,Bhattacharyya:2014uxa, Ghosh:2014vja, Ghosh:2014zra,
 Bhattacharyya:2015kda,Bhattacharyya:2016jsn}, 
%%%%%%%%%%%%%%%%%%%%%%%%%%%%%%%%%%%%%%%%%%%%%%%%%%%%%%%%%%%%%%%%%%%%%%%%
\begin{widetext}
%\begin{multline}
\begin{align}
\Omega^{\rm{Poly}}_{\mathrm{PNJL}}(\Phi,\bar{\Phi},
\sigma)=&\frac{\sigma^2}{2G}
-6N_f\int_0^\Lambda \varepsilon_p-2N_f T\int\frac{d^3p}
{(2\pi)^3}\text{ln}
[1+e^{-3\beta \varepsilon_q} +  N_c(\Phi+\bar{\Phi}e^{-
\beta \varepsilon_q})e^{-\beta \varepsilon_q}]\nonumber\\
&-2N_f T\int\frac{d^3p} {(2\pi)^3}  \text{ln}\left[1+e^{-3\beta 
\varepsilon_{\bar{q}}}
+N_c(\bar{\Phi}+\Phi e^{-\beta \varepsilon_{\bar{q}}}) 
e^{-\beta \varepsilon_{\bar{q}}}\right]+
U(\Phi,\bar{\Phi}) \nonumber \\
=&\Omega_\mathrm{q}+U(\Phi,\bar{\Phi})~,
	\label{eqomegaPNJL}
\end{align}

%\end{multline}
\end{widetext}
%%%%%%%%%%%%%%%%%%%%%%%%%%%%%%%%%%%%%%%%%%%%%%%%%%%%%%%%%%%%%%%%%%%%%%%%%
where $\beta$ is the inverse temperature, $N_f=2$ corresponds to the number of flavors and $\varepsilon_q=\varepsilon_{\bar{q}}=\sqrt{p^2+m^{*2}}$, $m^*$ being the constituent quark mass, given as, $m^*=m_0-G\langle\bar{\psi} \psi\rangle=m_0+\sigma$. The bare quark mass ($m_0$), the coupling strength ($G$) and the three momentum cutoff ($\Lambda$) are the model parameters which can be evaluated from three physical observables, the pion mass ($m_\pi$) the chiral condensate ($\langle\bar{\psi}\psi\rangle$), and the pion decay constant ($f_\pi$). In this work, we employ the parameter set as given in Ref.~\cite{PhysRevD.73.014019}.

Using the saddle point approximation, in the thermodynamic potential, one can replace $\Phi$, $\bar{\Phi}$ and $\sigma$ by their mean values, satisfying the saddle point equations,
%%%%%%%%%%%%%%%%%%%%%%%%%%%%%%%%%%%%%%%%%%%%%%%%%%%%%%%%%%%%%%%%%%%%%%%%%
\begin{eqnarray}
\left.\frac{\partial \Omega}{\partial \sigma} 
\right\rvert_{\sigma=\sigma_{\mathrm{mf}}}=0 ~;~ 
\left.\frac{\partial \Omega}{\partial \Phi}
\right\rvert_{\Phi=\Phi_{\mathrm{mf}}} =0 ~;~ 
\left.\frac{\partial \Omega}{\partial \bar{\Phi}}
\right\rvert_{\bar{\Phi}=\bar{\Phi}_{\mathrm{mf}}}\hspace{-2mm}=0~.
\label{saddle}
\end{eqnarray} 
%%%%%%%%%%%%%%%%%%%%%%%%%%%%%%%%%%%%%%%%%%%%%%%%%%%%%%%%%%%%%%%%%%%%%%%%%
With these replacements, thermodynamic observables can be evaluated using the known standard thermodynamic relations. 

%*************************************************************************
\subsubsection{\label{sssec:alt_appr} Quasiparticle approach}
%*****************************************************************************
Our proposal in this work is to couple the gluon quasi-particle 
model~\cite{Ruggieri:2012ny, Alba:2014lda,Islam:2012kv,Sasaki:2012bi,Islam:2021qwh} to the 2 flavour NJL model. In the usual saddle point approximation the thermodynamic potential is given as,
%####################################################################
\begin{align}
 \Omega^{\rm{qp}}_{\mathrm{PNJL}}=\Omega_\mathrm{q}+
 \Omega_{\mathrm{qp}}~,   
\end{align}
%###################################################################
where, $\Omega_\mathrm{q}$ is given in Eq.\eqref{eqomegaPNJL}, and $\Omega_\mathrm{qp}$ is given in Eq.~\eqref{eqquasi}, with $\Omega_{\mathrm{gqp}}$ given in Eq.\eqref{eqsimpleg}. This model can  provide a microscopic description for both quark and gluonic degrees of freedom, coupled to the background Polyakov field. However, with the usual saddle point approach we shall be stuck with the physical inconsistencies below the cross-over temperature as discussed earlier. Therefore in this work we adopt the proposed approach of integration over all possible $\theta_1$, $\theta_2$ configurations. 

We start from the thermodynamic potential given as,
%%%%%%%%%%%%%%%%%%%%%%%%%%%%%%%%%%%%%%%%%%%%%%%%%%%%%%%%%%
\begin{align}
    \Omega_{\rm PNJL}[\theta_1({\bf x}),\theta_2({\bf x}),
    \sigma({\bf x})]=&\Omega_{\mathrm{gqp}}[\theta_1({\bf x}),
    \theta_2({\bf x})]+\nonumber\\&~~~~\Omega_\mathrm{q}
    [\theta_1({\bf x}),\theta_2({\bf x}),\sigma({\bf x})].
\end{align}
%%%%%%%%%%%%%%%%%%%%%%%%%%%%%%%%%%%%%%%%%%%%%%%%%%%%%%%%%%
To evaluate the thermodynamic quantities we need to have an estimation of three independent fields, $\Phi,\bar{\Phi}$ and $\sigma$ or equivalently $\theta_1,\theta_2$ and $\sigma$. For $\theta_1$ and $\theta_2$ we shall implement the approach similar to that of the pure gauge study~\cite{Islam:2021qwh,Singha:2021wpi}. One could also adopt a similar strategy for the $\sigma$ field. But since the quark sector does not pose any issues like the gluon sector, and also for simplicity we shall continue to use the standard saddle point approximation of the thermodynamic potential with respect to the $\sigma$ field. We can then write the partition function for the 2-flavour QCD system, within our effective model framework,
replacing, $\Omega[\theta_1({\bf x}),\theta_2({\bf 
x}),\sigma({\bf x})]\rightarrow\Omega[\theta_1({\bf x}),\theta_2({\bf x}),\sigma]$, as,
%%%%%%%%%%%%%%%%%%%%%%%%%%%%%%%%%%%%%%%%%%%%%%%%%%%%%%%%%%
\begin{align}
  &Z_{\rm PNJL}[\sigma] \nonumber \\ &=\int{\mathcal{D}\theta_1\mathcal{D}\theta_2 
\mathrm{exp}\Bigg[-\frac{1}{T}\int 
d^3x\Omega_{\rm PNJL}[\theta_1({\bf x}),\theta_2({\bf x}),
\sigma] \Bigg]}
\label{eqpart}
\\
&= \int\prod_{\bf x} \frac{1}{24\pi^2}
d\theta_1({\bf x}) d\theta_2({\bf x}) \mathrm{Det}
_{\mathrm{VDM}}
\nonumber \\
&~~~~~~~~~\mathrm{exp}\left[-\frac{1}{T}\int d^3x 
\Omega_{\rm PNJL}[\theta_1({\bf x}),\theta_2({\bf x}),
\sigma]\right].
\label{eqpointpart}
\end{align}
%%%%%%%%%%%%%%%%%%%%%%%%%%%%%%%%%%%%%%%%%%%%%%%%%%%%%%%%%%
\noindent
Considering constant field configurations the configuration space may be broken up into $\mathrm{N}\rightarrow \infty$ equivalent and independent points. The partition function may be written as,
%%%%%%%%%%%%%%%%%%%%%%%%%%%%%%%%%%%%%%%%%%%%%%%%%%%%%%%%%%
\begin{align}
    Z_{\rm PNJL}[\sigma]=z^N_{\rm PNJL}[\sigma],
\end{align}
%%%%%%%%%%%%%%%%%%%%%%%%%%%%%%%%%%%%%%%%%%%%%%%%%%%%%%%%%%
where,
%%%%%%%%%%%%%%%%%%%%%%%%%%%%%%%%%%%%%%%%%%%%%%%%%%%%%%%%%%
\begin{eqnarray}
z_{\rm PNJL}[\sigma]\ &=& \int \frac{1}{24\pi^2}d\theta_1 d\theta_2
\mathrm{Det}_{\mathrm{VDM}} \nonumber \\
&&~~~~~~~~~ \mathrm{exp}\left[-\frac{v}{T}
\omega_{\rm PNJL}[\theta_1,\theta_2,\sigma]\right].
\label{eqsmallqpart}
\end{eqnarray}
%%%%%%%%%%%%%%%%%%%%%%%%%%%%%%%%%%%%%%%%%%%%%%%%%%%%%%%%%%
\noindent
While the $\theta$ fields are integrated out, the $\sigma_{\rm mf}$ may be obtained from the saddle point of the effective thermodynamic potential $\omega_{\rm PNJL}=-(T/v)\ln z_{\rm PNJL}$, 
%%%%%%%%%%%%%%%%%%%%%%%%%%%%%%%%%%%%%%%%%%%%%%%%%%%%%%%%%%
\begin{align}
    \left.\frac{\partial \omega_{\rm PNJL}}{\partial \sigma}\right|_{\sigma=\sigma_\mathrm{mf}}=0~.
    \label{eqsadalter}
\end{align}
%%%%%%%%%%%%%%%%%%%%%%%%%%%%%%%%%%%%%%%%%%%%%%%%%%%%%%%%%%
 Obtaining the solution, $\sigma_{\rm mf}$, the thermodynamic potential $\omega_{\rm PNJL}[\sigma_{\rm mf}]$ can be evaluated. Thereafter,
we can compute the thermodynamic quantities following the 
usual prescription,
%%%%%%%%%%%%%%%%%%%%%%%%%%%%%%%%%%%%%%%%%%%%%%%%%%%%%%%%%%
\begin{align}
    p&=-\omega_{\rm PNJL}[\sigma_{\rm mf}],\label{eqpresf}\\
    s&=\frac{\partial p}{\partial T},\label{eqentf}\\
    \epsilon&=sT-p,\label{eqenerf}\\
    C_V&=\frac{\partial \epsilon}{\partial T}\,\, \mathrm{and}\label{eqcvf}
    \\
    v_s^2&=\frac{\partial p}{\partial \epsilon}
    =\frac{\partial p}{\partial T}/\frac{\partial 
    \epsilon}{\partial T}=\frac{s}{C_V}\label{eqvelf}~.
\end{align}
%%%%%%%%%%%%%%%%%%%%%%%%%%%%%%%%%%%%%%%%%%%%%%%%%%%%%%%%%%
Further note that the expectation values of any $\Phi$, $\bar \Phi$ or $\theta_1$, $\theta_2$ dependent observables can be obtained as,
%%%%%%%%%%%%%%%%%%%%%%%%%%%%%%%%%%%%%%%%%%%%%%%%%%%%%%%%%%
\begin{align}
<O[\Phi,\bar\Phi]> =& \frac{1}{z}\int \frac{1}{24\pi^2}
d\theta_1 d\theta_2
\mathrm{Det}_{\mathrm{VDM}}
 O[\Phi,\bar\Phi]\nonumber \\
&~~~\mathrm{exp}\left[-\frac{v}{T}
\omega[\theta_1,\theta_2,\sigma_\mathrm{mf}]\right]~.
\label{eqexpect2}
\end{align}
%%%%%%%%%%%%%%%%%%%%%%%%%%%%%%%%%%%%%%%%%%%%%%%%%%%%%%%%%%
\noindent
For a numerical estimation of these expectation values and the thermodynamic quantities we need to specify our model parameters. Following the same prescription as in our previous pure gauge study, we will introduce a quasi-gluon mass as a phenomenological parameter and replace $|\vec{p}|$ in Eq.~\eqref{eqsimpleg} by $\varepsilon_g=\sqrt{|\vec{p}| ^2+m_g^2}$.  So in the current approach, along with the usual NJL parameters we have an additional parameter $v$ and the effective quasi-gluon mass. 

\section{\label{sec:result} Result}
%%%%%%%%%%%%%%%%%%%%%%%%%%%%%%%%%%%%%%%%%%%%%%%%%%%%%%%%%%%%%%%%%%%%%%%%%%%%
%***********************************************************************
\subsection{\label{ssec:param}Model parameters}
%***********************************************************************
The NJL model has 3 parameters as already mentioned in the subsection.\ref{sssec:std_appr}. In Tab.~\ref{table}, we show their values as taken from Ref.~\cite{PhysRevD.73.014019} and used in our numerical analysis.
%%%%%%%%%%%%%%%%%%%%%%%%%%%%%%%%%%%%%%%%%%%%%%%%%%%%%%%%%%
\begin{table}
	\centering
	\begin{tabular}{|c|c|c|}
		\hline
		$\Lambda$ (GeV)&$m_0$(GeV) & G($\text{GeV}^{-2}$)
		\\[1 ex]
		%		\hline
		\hline
		0.651 & 0.0055 & 10.08\\[1 ex]
		\hline
	\end{tabular}
	\caption{Parameters for NJL model.}
	\label{table}
\end{table}
%%%%%%%%%%%%%%%%%%%%%%%%%%%%%%%%%%%%%%%%%%%%%%%%%%%%%%%%%%
\noindent
From the quasi-gluon sector, we need to fix the model parameter $v$ 
and we choose it to be the same as in our pure gauge study~\cite{Islam:2021qwh} i.e. $v=(0.5 T_d)^{-3}$. 
In Ref.~\cite{Islam:2021qwh}, we kept the value of the deconfinement temperature, $T_d$, arbitrary in the numerical simulations and expressed our results in terms of the scaled temperature, $T/T_d$. In our present study, however, we expect the transition temperature to be modified due to the presence of the quark degrees of freedom and the 
additional chiral dynamics in the system. So in this work we will treat $T_d$ as an independent free parameter that can be tuned to obtain the qualitative agreement to the lattice results. In the following  we took  $T_d = 270$ MeV to report our observation.

\begin{figure}
	{\includegraphics[width=\linewidth]
	{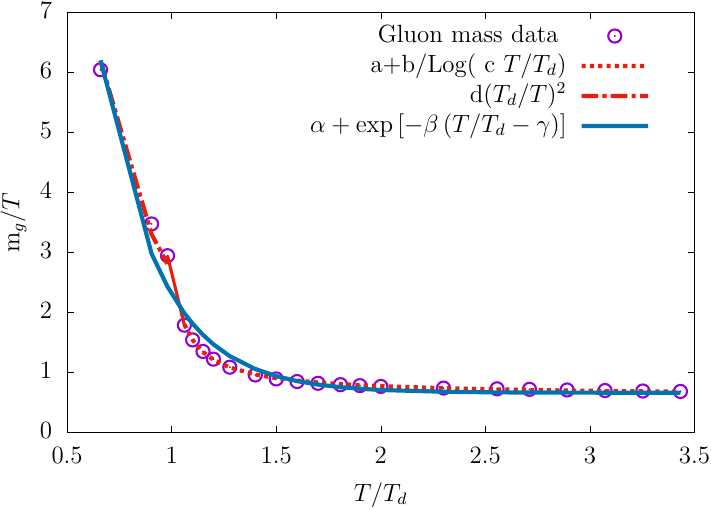}} 
	\caption{Effective gluon mass as function of temperature fitted to the data~\cite{Islam:2021qwh} extracted from the pressure obtained in Lattice Yang-Mills theory~\cite{Giusti:2016iqr}.}
	\label{fg.mass}
\end{figure}

In Ref.~\cite{Islam:2021qwh}, we introduced an effective mass parameter for the quasi-gluons to capture the confinement physics 
on a quantitative level. Comparing our result with the lattice pure gauge pressure at the continuum limit~\cite{Giusti:2016iqr}, we 
extracted the value of the mass parameter at different temperatures, which  are shown in Fig.\ref{fg.mass} as ``gluon mass  data''. For a pure gauge system, the  confinement-deconfinemt  transition is of first order and as a  consequence, one can observe a sharp change in 
 the extracted data points at $T/T_d=1$. To mimic  that scenario, in Ref~\cite{Islam:2021qwh}, we  took two different analytic form of the thermal gluon  mass at two different  temperature range as the following,
%%%%%%%%%%%%%%%%%%%%%%%%%%%%%%%%%%%%%%%%%%%%%%%%
 \begin{eqnarray}
m_g(T)/T &=& a+ b/\ln(c~T/T_d), \rm{~for~} T/T_d > 1 
\\
         &=& d~(T_d/T)^2, \rm{~for~} T/T_d < 1 .
\label{gluon_mass}
\end{eqnarray}
 %%%%%%%%%%%%%%%%%%%%%%%%%%%%%%%%%%%%%%%%%%%%%%%%%%%%%%%%%
 The fitted values of the parameters can be found in Ref.~\cite{Islam:2021qwh}. 
 
In the presence of the quark degrees of freedom, the centre symmetry is explicitly broken. Consequently, the phase transition changes to a crossover and the discontinuities get smoothed out. For a two-flavour system, to incorporate the qualitative nature of a crossover transition with a minimum deviation from the pure gauge model, we have used a continuous parametric form of effective gluon mass,
%%%%%%%%%%%%%%%%%%%%%%%%%%%%%%%%%%%%%%%%%%%%%%%%%%%%%%%%%%
\begin{align}
      m_g(T)/T=\alpha+\mathrm{exp}\left[-
      \beta\left(\frac{T}{T_d}-\gamma\right)\right]~,
\end{align}
%%%%%%%%%%%%%%%%%%%%%%%%%%%%%%%%%%%%%%%%%%%%%%%%%%%%%%%%%%
where, the parameters $\alpha,\beta,\gamma$ are fitted to the same mass data obtained from the pressure of the pure gauge system~\cite{Giusti:2016iqr}. The extracted values of the parameters, evaluated by least square fit, are given in the Tab.~\ref{tablemass}. In the numerical analysis the central values of the parameters are taken.
%%%%%%%%%%%%%%%%%%%%%%%%%%%%%%%%%%%%%%%%%%%%%%%%%%%%%%%%%%
\begin{table}
\begin{center}
	\begin{tabular}{|c|c|c|c|c|}
		\hline
		\multicolumn{3}{|c|}{Fitted}& \multicolumn{2}{|c|}
		{Chosen}\\
		\hline
		& \\[-3.3 ex]
$\alpha$ & $\beta$ & $\gamma$ & $T_d$(GeV) & $v$(GeV$^{-3}
$)\\[1 ex]
		\hline
		& \\[-3.3 ex]
		0.662899 & 3.56978 & 1.13963 & 0.270  & 
		$(0.5T_d)^{-3}$ \\[1 ex]
		$\pm 0.0652$ & $\pm 0.2224$ & $\pm 0.02875$ &  & \\[1 ex]
		\hline
	\end{tabular}
	\caption{Parameters for quasi-gluon sector.}
	\label{tablemass}
	\end{center}
\end{table}
%%%%%%%%%%%%%%%%%%%%%%%%%%%%%%%%%%%%%%%%%%%%%%%%%%%%%%%%%%
\noindent
In Fig.\ref{fg.mass}, along with the gluon mass data, we have shown the smooth parametric function used as quasi-gluon mass in the 
current study and two analytic forms of the effective gluon mass used for our pure gauge study~\cite{Islam:2021qwh}. \\
With these model parameters we are now ready to discuss the order parameters and the medium thermodynamics obtained within our model set up.
%%%%%%%%%%%%%%%%%%%%%%%%%%%%%%%%%%%%%%%%%%%%%%%%%%%%%%%%%%
%%%%%%%%%%%%%%%%%%%%%%%%%%%%%%%%%%%%%%%%%%%%%%%%%%%%%%%%%%

%**********************************************************************
\subsection{\label{ssec:OP}Order parameters}
%**********************************************************************
%%%%%%%%%%%%%%%%%%%%%%%%%%%%%%%%%%%%%%%%%%%%%%%%%%%%%%%%%%
\begin{figure}
	{\includegraphics[width=\linewidth]
	{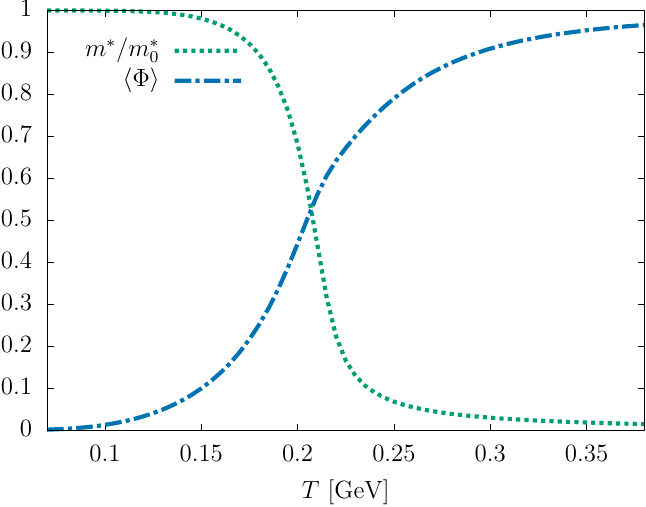}} 
	\caption{Thermal evolution of the constituent quark mass and the expectation value of the Polyakov loop, $\langle \Phi\rangle$.}
	\label{fg.sigma_phi}
\end{figure} 
%%%%%%%%%%%%%%%%%%%%%%%%%%%%%%%%%%%%%%%%%%%%%%%%%%%%%%%%%%
For the chiral transition the expectation value of chiral condensate, $\sigma_{\rm mf}$ is obtained using the saddle point approximation from Eq.~\eqref{eqsadalter}. In Fig.\ref{fg.sigma_phi} we have shown the temperature variation of the constituent quark mass normalized by its zero temperature value, given as, $m^*/m^*_0=\frac{m_0+\sigma_{\rm{mf}}}{m_0+\sigma_{\rm{mf}}(T=0)}$ . The order parameter for the confinement-deconfinement transition, $\langle \Phi\rangle$, is obtained From Eq.~\eqref{eqexpect2} and shown in Fig.\ref{fg.sigma_phi} with blue dashed-dot line. Both the order parameters vary smoothly with temperature demonstrating a crossover scenario.  The nonzero chiral condensate gradually goes to zero with increasing temperature as the chiral symmetry is restored. On the other hand, $\langle\Phi\rangle=0$ describes the confined phase, that slowly increases to 1 with increasing temperature and the system becomes deconfined.  
In Ref.~\cite{Islam:2021qwh}, 
we observed that $\langle \Phi \rangle$ obtained from Eq.~\eqref{eqexpect2} for a pure gauge system, remains zero even for deconfined  Z(3) symmetry broken phase. We pointed out that without an explicit symmetry breaking term, all the three degenerate Z(3) ground states are equally likely, and within our prescription that leads the expectation value to zero. 
In the present case the dynamical quarks break the center symmetry 
explicitly~\cite{confinement:2011bk,Hasenfratz:1983ce}. 
As a consequence, the ground state with the real value of $\Phi$ is preferred at the high temperature , and we obtain a nontrivial expectation value.  
%%%%%%%%%%%%%%%%%%%%%%%%%%%%%%%%%%%%%%%%%%%%%%%%%%%%%%%%%%
\begin{figure}
{\includegraphics[width=\linewidth]
{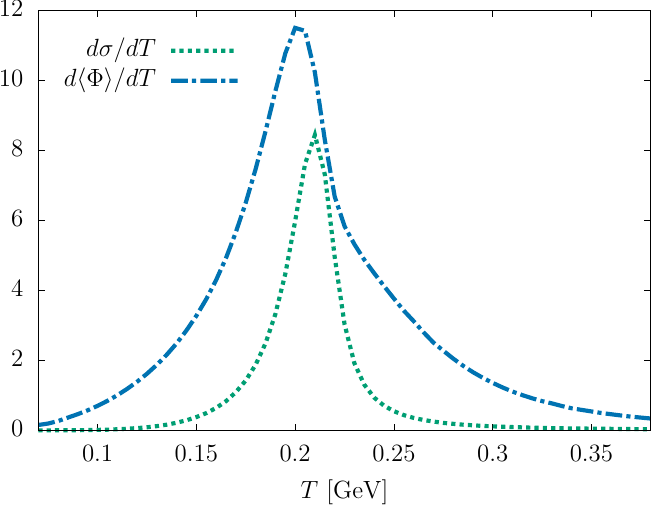}} 
	\caption{Thermal evolution of the temperature derivatives $d\langle \Phi\rangle/dT$ and $d\sigma/dT$.}
	\label{fig:critical}
\end{figure}
%%%%%%%%%%%%%%%%%%%%%%%%%%%%%%%%%%%%%%%%%%%%%%%%%%%%%%%%%%
For an estimate of the crossover temperature, we obtain the temperature derivative of the order parameters as shown in Fig.\ref{fig:critical}. The peaks representing the corresponding crossover temperatures, can be seen to lie close to each other and their average is used as the crossover temperature $T_c \approx 0.2065$ GeV in our remaining numerical analysis. This $T_c$ is in good agreement with the corresponding values reported on the lattice~\cite{Burger:2014xga}.
%**********************************************************************
\subsection{\label{ssec:Therm}Thermodynamic observables } 
%**********************************************************************

%%%%%%%%%%%%%%%%%%%%%%%%%%%%%%%%%%%%%%%%%%%%%%%%%%%%%%%%%%
\begin{figure}
{\includegraphics[width=\linewidth]{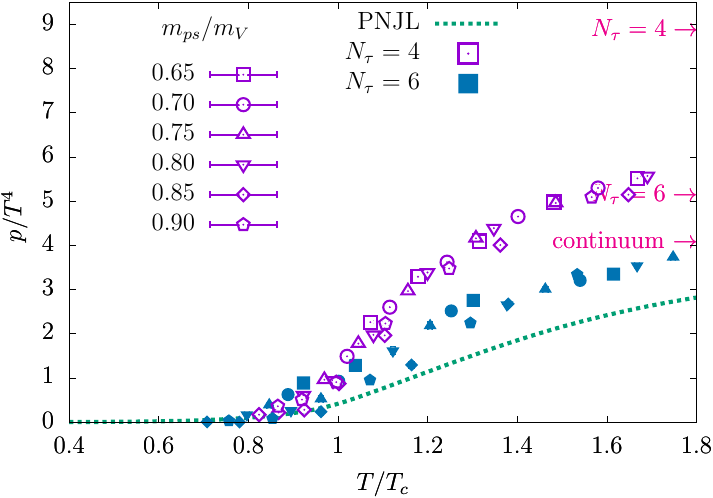}} 
	\caption{Scaled pressure as a function of temperature. Lattice data for $N_\tau=4$ (violet open symbols) and $N_\tau=6$ (blue solid symbols) are taken from Ref.~\cite{CP-PACS:2001hxw} for various $m_{ps}/m_V$ ratios. The Stefan-Boltzmann limit for $N_\tau=4,N_\tau=6$ and continuum limit are shown by magenta arrows.}
	\label{fg.pressure}
\end{figure}
%%%%%%%%%%%%%%%%%%%%%%%%%%%%%%%%%%%%%%%%%%%%%%%%%%%%%%%%%%

In Fig.~\ref{fg.pressure} we show the thermal variation of the pressure scaled with $T^4$. Unlike the pure gauge system~\cite{Islam:2021qwh}, here we observe a smooth increase in pressure near the transition temperature, as expected for a crossover. For comparison we plot the available 2-flavor Lattice QCD data for finite ($N_\tau=4$ and $N_\tau=6$) lattices with various fixed pseudo scalar to vector meson mass ratios, $m_{ps}/m_V$ ~\cite{CP-PACS:2001hxw} which are shown in the plot with different symbols. We have used the open symbols for $N_\tau =4$ and solid symbols for $N_\tau=6$.  The corresponding Stefan-Boltzmann limits are indicated on the plot by arrows. Our results are qualitatively consistent with the Lattice results. Plots for continuum values of scaled pressure measured on the lattice with twisted mass Wilson quarks are available in Ref.~\cite{Burger:2014xga}. It reaches only about half of the Stefan-Boltzmann limit at around 2$T_c$. This could be because of heavier quark masses used in the simulation. 

%%%%%%%%%%%%%%%%%%%%%%%%%%%%%%%%%%%%%%%%%%%%%%%%%%%%%%%%%%
%%%%%%%%%%%%%%
\begin{figure}
{\includegraphics[width=\linewidth]{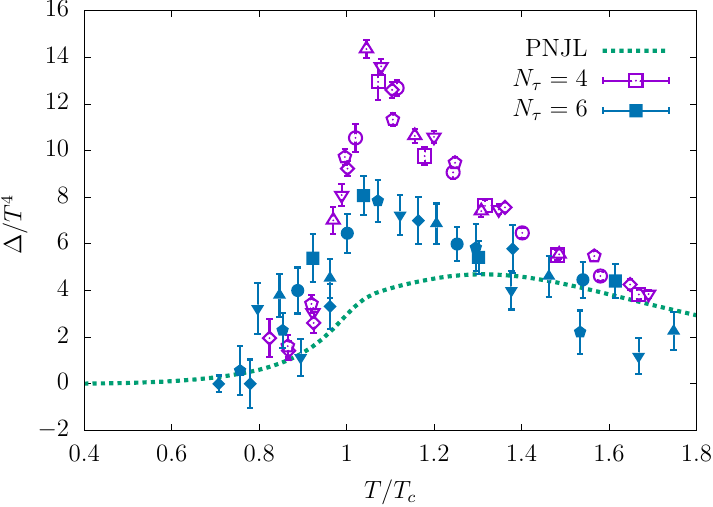}}
%\\ {\includegraphics[width=\linewidth]{images/intbyen.pdf}}
\caption{Temperature variations of scaled conformal measure 
%and (b) conformal measure to energy density ratio. 
(lattice data from~\cite{CP-PACS:2001hxw} and symbols are same as Fig.\ref{fg.pressure}).}
\label{fg.conformal}
\end{figure} 
%%%%%%%%%%%%%%%%%%%%%%%%%%%%%%%%%%%%%%%%%%%%%%%%%%%%%%%%%%
In Fig.~\ref{fg.conformal}(a) the interaction measure, $\Delta=\epsilon-3p$, scaled with $T^4$ is shown as a function of $T/T_c$. The corresponding lattice QCD data are also presented. We note that with increasing $N_\tau$ the peak in the interaction measure gradually flattens, that agrees with our result. However above $T_c$ the decrease in the interaction measure is noticeably slower, an effect to which we will return shortly.
%%%%%%%%%%%%%%%%%%%%%%%%%%%%%%%%%%%%%%%%%%%%%%%%%%%%%%%%%%
\begin{figure}{\includegraphics[width=\linewidth]
{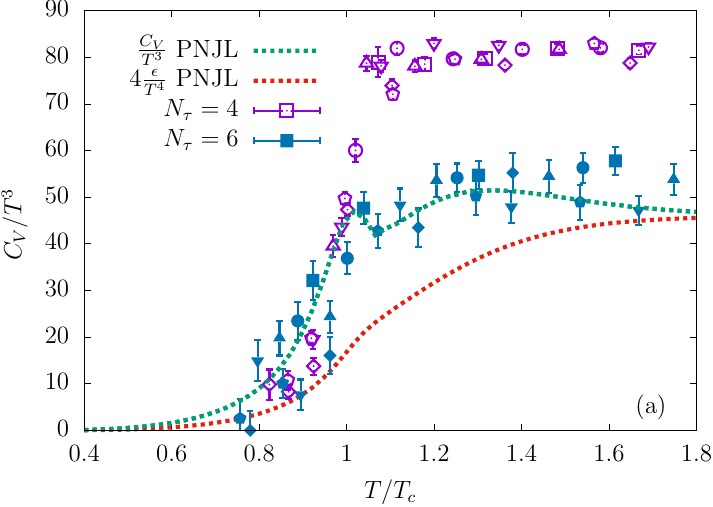}}\\
{\includegraphics[width=\linewidth]{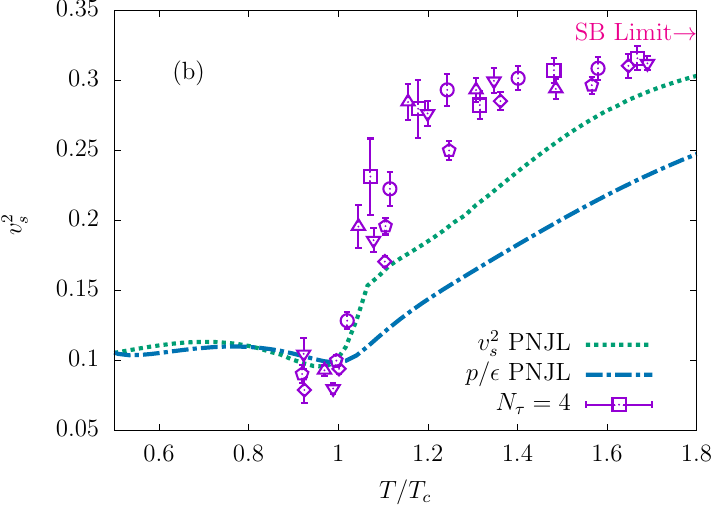}}
\caption{Temperature variation of (a) scaled 
specific heat 
and (b) speed of sound square.
 (lattice data for $4\epsilon/T^4$ and $v_s^2$ are from Ref.~\cite{CP-PACS:2001hxw}).}
\label{fg.cvcs2}
\end{figure} 
%%%%%%%%%%%%%%%%%%%%%%%%%%%%%%%%%%%%%%%%%%%%%%%%%%%%%%%%%% 

The scaled specific heat, $C_V/T^3$ and four times the scaled energy density, $4\epsilon/T^4$, are presented in Fig.~\ref{fg.cvcs2}(a). For 
a conformal theory, $C_V/T^3=4\epsilon/T^4$. The difference between the thermal behavior of these two quantities is another measure of deviation from the conformal limit. The bulge in specific heat 
at a temperature above $T_c$ is a consequence of the flattened peak in the interaction measure above $T_c$~\ref{fg.conformal}(a). This is because the specific heat is related to the temperature derivative of the interaction measure~\cite{HotQCD:2014kol}. The lattice results shown are for $4\epsilon/T^4$ obtained for $N_\tau=4$ and $N_\tau=6$.

In Fig.~\ref{fg.cvcs2}(b) the squared speed of sound is shown along with $p/\epsilon$. The available lattice data of squared speed of sound  is for  $N_\tau=4$ lattice~\cite{CP-PACS:2001hxw}. The minima of the speed of sound coincides with the point where the pressure to energy density ratio is minimum or the softest  point of the equation of state~\cite{Hung:1994eq}. However somewhat above $T_c$ the curve shows a shallow dip which corresponds to the bulge in the specific heat mentioned above. Below we try to explain the source of these observed features. 

%%%%%%%%%%%%%%%%%%%%%%%%%%%%%%%%%%%%%%%%%%%%%%%%%%%%%%%%%%
\begin{figure}
{\includegraphics[width=\linewidth]{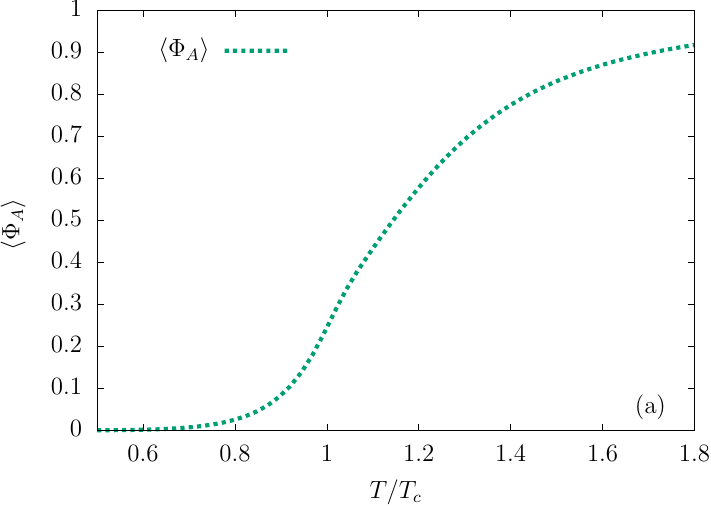}}\\
{\includegraphics[width=\linewidth]{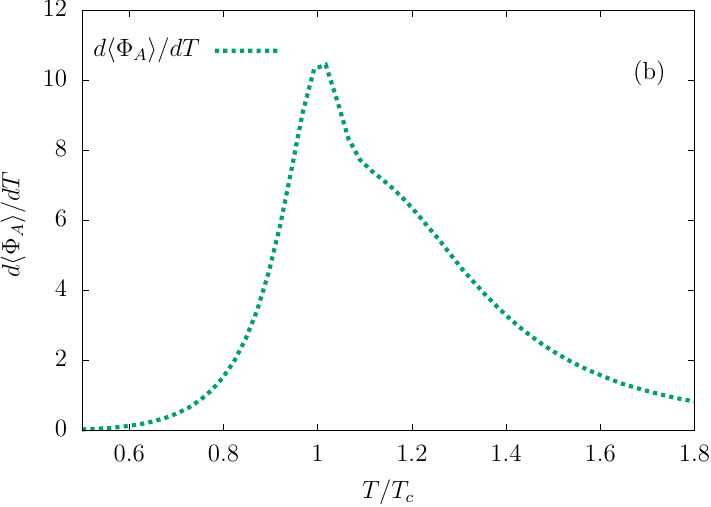}}
\caption{Thermal evolution of (a) $\langle\Phi_A\rangle$ and (b) $d\langle\Phi_A\rangle/dT$.}
\label{fg.adj}
\end{figure} 
%%%%%%%%%%%%%%%%%%%%%%%%%%%%%%%%%%%%%%%%%%%%%%%%%%%%%%%%%%

The thermal average of the adjoint Polyakov loop $\langle\Phi_A\rangle$ is related to the free energy of the quasigluons~\cite{Abramchuk:2018bco, Simonov:2010bf, Marsh:2013xsa, Gupta:2007ax, etde_20896960}.
Its temperature variation is shown in  Fig.~\ref{fg.adj}(a). In Fig.~\ref{fg.adj}(b) the thermal derivative of $\langle\Phi_A\rangle$ is presented. 
We note that there are two factors deciding thermal variation of $\frac{d\langle\Phi_A\rangle}{dT}$. First one is the coupling of the gluonic sector to the quark sector leading to the prominent peak near $T_c$. The second one is the model input in the gluonic sector itself through the parameter $T_d=270$ MeV. This is the factor that leads to the small bulge at a temperature just above $T_c$. Since $T_d$ controls the thermal behavior of $m_g$, one can say that in our model while the chiral crossover and quark deconfinement occurs very close to each other near $T_c$, the deconfinement of gluons occur at a higher temperature. This is what gives rise to the bulge in the specific heat and dip in speed of sound as discussed above.

%%%%%%%%%%%%%%%%%%%%%%%%%%%%%%%%%%%%%%%%%%%%%%%%%%%%%%%%%%
\begin{figure}
    \centering
   \includegraphics[width=\linewidth]{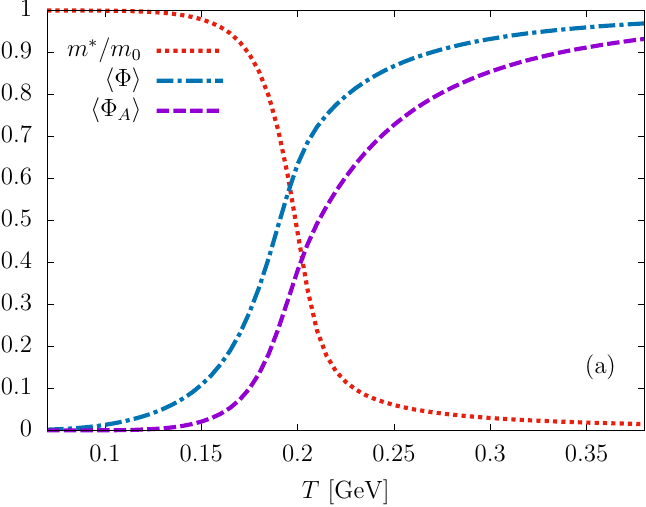} \\\includegraphics[width=\linewidth]{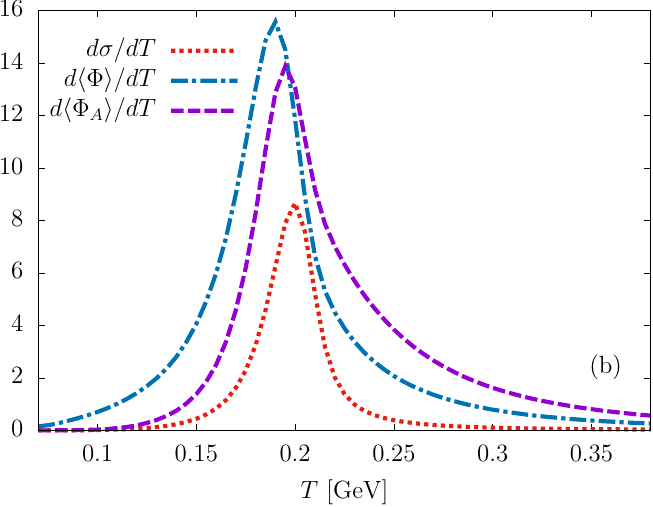}
    \caption{The order parameters (a) and their thermal derivatives (b) with $T_d=0.170$ GeV. }
    \label{fig:placeholder}
\end{figure}
%%%%%%%%%%%%%%%%%%%%%%%%%%%%%%%%%%%%%%%%%%%%%%%%%%%%%%%%%%

%%%%%%%%%%%%%%%%%%%%%%%%%%%%%%%%%%%%%%%%%%%%%%%%%%%%%%%%%%
\begin{figure}{\includegraphics[width=\linewidth]
{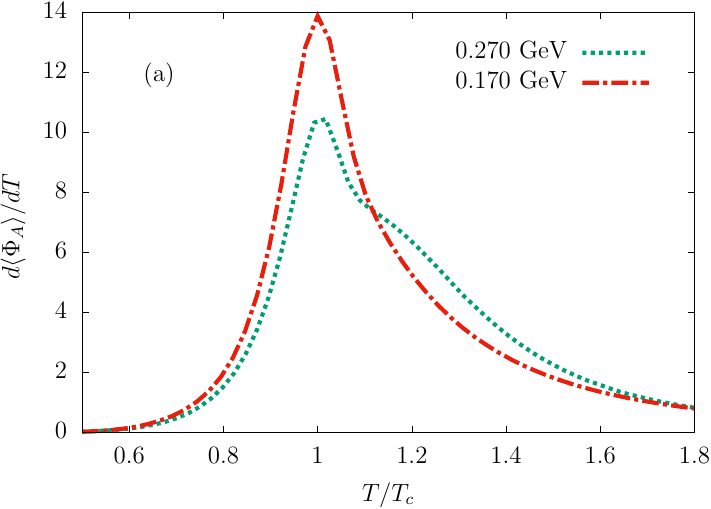}}\\
{\includegraphics[width=\linewidth]
{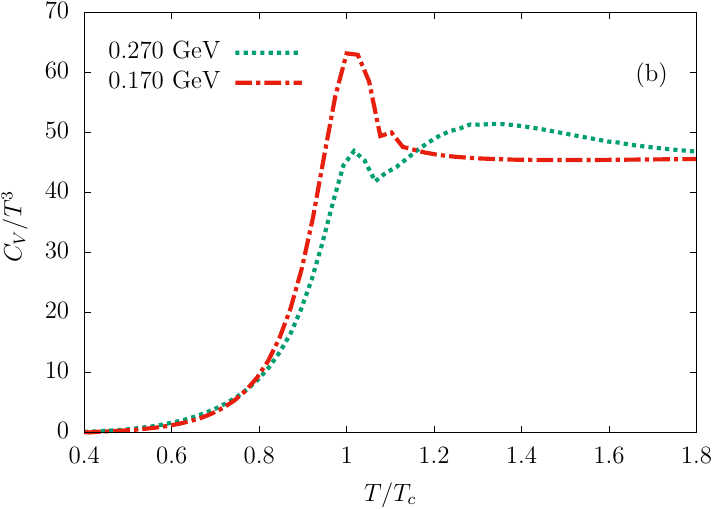}}	\\
{\includegraphics[width=\linewidth]
{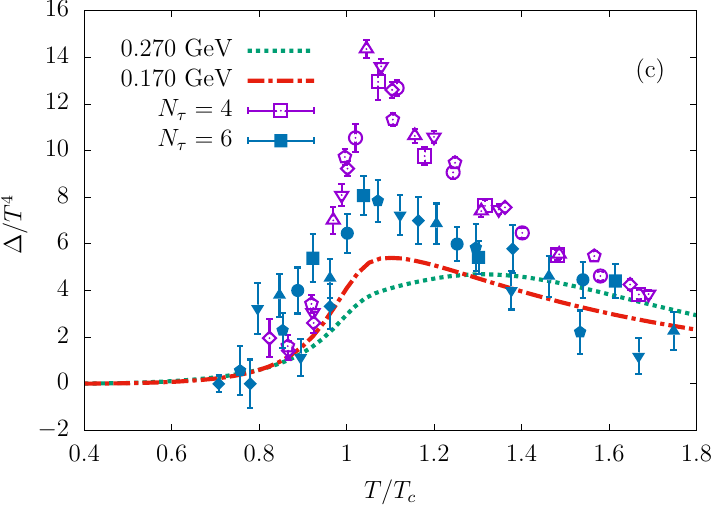}}
\caption{Comparative study of (a) thermal 
derivative of adjoint Polyakov loop, (b) specific 
heat and (c) interaction measure for two 
different values of mass parameter, $T_d$}
\label{fg.comp}
\end{figure} 
%%%%%%%%%%%%%%%%%%%%%%%%%%%%%%%%%%%%%%%%%%%%%%%%%%%%%%%%%%

Since we kept $T_d$ as a free parameter of our model set up,  we can fine tune it to make our model more phenomenologically consistent. The chiral and deconfinement transitions come close to each other by reducing the value of $T_d$ from $0.27$ GeV to $0.17$ GeV, as was done in some of our earlier works (see e.g.~\cite{PhysRevD.73.014019}). This is shown in Fig.~\ref{fig:placeholder} through the order parameters and their derivatives.  We further obtain the comparative plots of various thermodynamic observables as function of temperature for the two different values of $T_d$ as shown in Fig.~\ref{fg.comp}. As expected there is no secondary features above $T_c$ either for 
$\frac{d\langle\Phi_A\rangle}{dT}$ or for $C_V/T^3$ and for $\Delta/T^4$. We shall therefore use $T_d=170$, keeping all the other parameter values mentioned earlier unchanged in the following discussions.

We have now completed building up the basic formalism for a quasi-particle description of strongly interacting matter within the two-flavor PNJL model. In the following we discuss one of its applications where transport coefficients are computed taking in contributions from both the quark and gluon sectors that was nonviable till date.

%------------------------------------------------------
\section{\label{sec:App} Distribution functions and transport coefficients}
%-----------------------------------------------------
The present quasi-particle PNJL framework enables us to obtain the  
thermal distribution functions for both quasi-quarks and quasi-gluons, which is a major advantage over the usual Polyakov extended chiral effective models. As an example, in this section we will discuss the estimation of the transport coefficients $-$ shear ($\eta$) and bulk ($\zeta$) viscosities of the QCD medium. These viscosities have already been studied in detail within various model frameworks. In NJL models the viscosities have been studied in Refs.~\cite{Sasaki:2008um, PhysRevC.88.068201, Marty:2013jga, Lang:2013lla, Lang:2013lla, Lang:2015nca, Deb:2016myz, Harutyunyan:2017ttz, Ghosh:2018xll}. In PNJL models similar studies have been reported in Refs.~\cite{PhysRevD.91.054005, Saha:2017xjq, Zhao:2020xob, yp8h-pdkr}. Several authors have studied the viscosities in other quark-meson models from various perspectives ~\cite{Chakraborty:2010fr, Tawfik:2016edq, Abhishek:2017pkp, Singha:2017jmq}. However in all these studies the quasi-particles that contribute to the viscosities were either quarks or mesons or both. On the other hand estimation of transport coefficients including both quark and gluon quasiparticles were done in Ref.~\cite{Mykhaylova:2019wci}, following the quasiparticle model discussed in Refs.~\cite{Bluhm:2004xn, Bluhm:2006yh, Bluhm:2007nu}. However there the quasi-particles follow the ideal distributions with masses fitted to the lattice data to mimic the chiral and confinement transitions. \\
In the following we discuss the distribution functions for quasi-gluons and quasi-quarks in subsection.\ref{ssec:distri} followed by an estimation of  transport coefficients in subsection\ref{ssec:trans}.
%%%%%%%%%%%%%%%%%%%%%%%%%%%%%%%%%%%%%%%%%%%%%%%%%%%%%%%%%%
%%%%%%%%%%%%%%%%%%%%%%%%%%%%%%%%%%%%%%%%%%%%%%%%%%%%%%%%%%
\begin{figure}
	{\includegraphics[width=\linewidth]{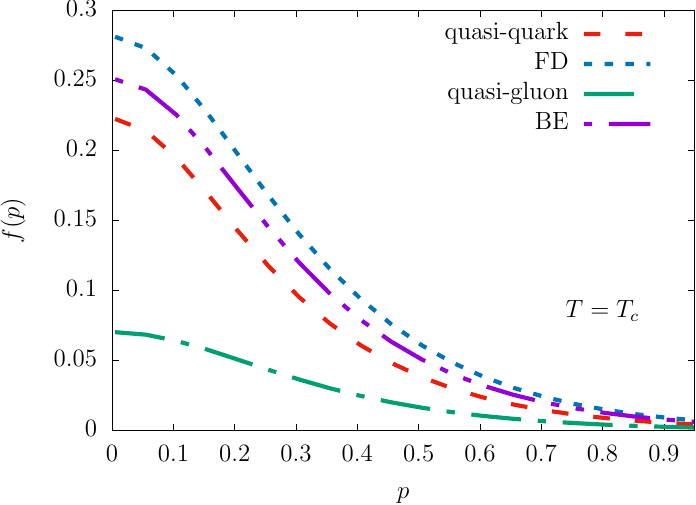}} 
	\caption{Momentum distributions of quark and 
 gluonic quasi-particles at temperature $T=T_c$.}
	\label{fg.distT}
\end{figure} 
%%%%%%%%%%%%%%%%%%%%%%%%%%%%%%%%%%%%%%%%%%%%%%%%
\subsection{\label{ssec:distri}Distribution function}
\begin{figure}
{\includegraphics[width=\linewidth]
{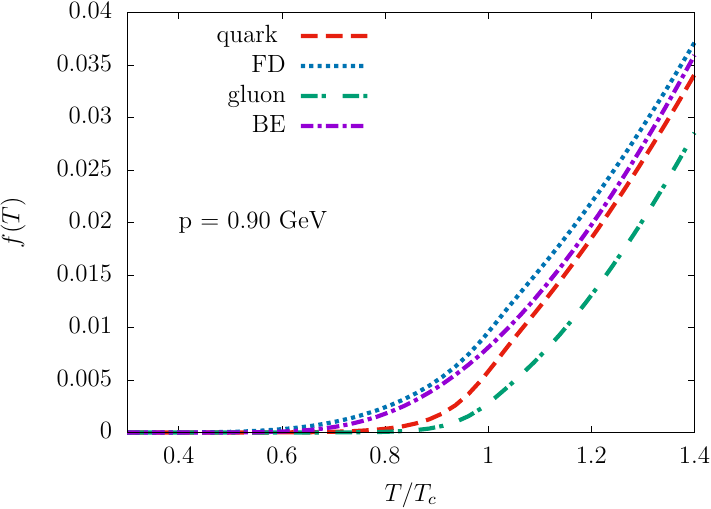}
}
\caption{Temperature distribution of  quark 
and  gluon quasi-particles for 
$|\vec{p}_1|=0.90$ GeV.}
\label{fg.dist}
\end{figure}

%-----------------------------------------------------
The medium modified color averaged distribution functions for gluon and quark quasi-particles are obtained from the partition function given in Eq.\eqref{eqsmallqpart}\cite{Hansen:2006ee}. For the gluon quasi-particles we have ,
%%%%%%%%%%%%%%%%%%%%%%%%%%%%%%%%%%%%%%%%%%%%%%%%%%%%%%%%%%
\begin{align}
f_g(p_1,T)&= -\frac{2}{d_g24\pi^2 z}\int d\theta_1 d\theta_2
\mathrm{Det}_{\mathrm{VDM}} \mathrm{e}^{-\frac{v}{T}
\omega[\theta_1,\theta_2,\sigma_\mathrm{mf}]}\nonumber
\\&\,~~ \times\frac{\sum\limits_{n=1}^8 
na_n\mathrm{exp}
\left[-\frac{n\varepsilon_g(p_1)}{T}\right]}{1+
\sum\limits^{8}_{n=1} a_n e^{-\frac{n \varepsilon_g(p_1)}
{T}}},
\label{eq.distri_gluon}
\end{align}
%%%%%%%%%%%%%%%%%%%%%%%%%%%%%%%%%%%%%%%%%%%%%%%%%%%%%%%%%%
and for quarks and anti-quarks we have, respectively,
%%%%%%%%%%%%%%%%%%%%%%%%%%%%%%%%%%%%%%%%%%%%%%%%%%%%%%%%%%
\begin{align}
    f_q(p_1,T)&=\frac{2 N_f 
    N_c}{d_q 24\pi^2 z}\int d\theta_1 d\theta_2 
    \mathrm{Det}_{\mathrm{VDM}} \mathrm{e}^{-\frac{v}{T}
\omega[\theta_1,\theta_2,\sigma_\mathrm{mf}]}\times 
\nonumber\\&
\frac{e^{-3\frac{ \varepsilon_q(p_1)}{T}}+
\left(\Phi+2\bar{\Phi} e^{-\frac{\varepsilon_q(p_1)}{T}}\right) e^{-
\frac{\varepsilon_q(p_1)}{T}}}{1+e^{-3\frac{\varepsilon_q(p_1)}{T}}
+N_c\left(\Phi+\bar{\Phi}e^{-\frac{\varepsilon_q(p_1)}{T}}\right)e^{-\frac{
\varepsilon_q(p_1)}{T}}}~,
 \label{eq.distri_quark}
\end{align}
%%%%%%%%%%%%%%%%%%%%%%%%%%%%%%%%%%%%%%%%%%%%%%%%%%%%%%%%%
and,
%%%%%%%%%%%%%%%%%%%%%%%%%%%%%%%%%%%%%%%%%%%%%%%%%%%%%%%%%%
\begin{align}
    f_{\bar{q}}(p_1,T)&=\frac{2 N_f N_c}{d_q 24\pi^2 z}\int 
    d\theta_1 d\theta_2 \mathrm{Det}_{\mathrm{VDM}} 
    \mathrm{e}^{-\frac{v}{T}
\omega[\theta_1,\theta_2,\sigma_\mathrm{mf}]}\times 
\nonumber\\&\frac{e^{-3\frac{\varepsilon_{\bar{q}}
(p_1)}{T}}+\left(\bar{\Phi}+2\Phi e^{-\frac{\varepsilon_{\bar{q}}
(p_1)}{T}}\right) e^{-\frac{\varepsilon_{\bar{q}}(p_1)}{T}}}
{1+e^{-3\frac{\varepsilon_{\bar{q}}(p_1)}{T}}+N_c\left(\bar{\Phi}+
\Phi e^{-\frac{\varepsilon_{\bar{q}}(p_1)}{T}}\right)e^{-
\frac{\varepsilon_{\bar{q}}(p_1)}{T}}}.
\label{eq.distri_quark_ant}
\end{align}
%%%%%%%%%%%%%%%%%%%%%%%%%%%%%%%%%%%%%%%%%%%%%%%%%%%%%%%%%
\noindent
Here $d_g$ is the gluon degeneracy factor and $d_q$ is the degeneracy 
factor for quarks or anti-quarks. In the present work,  we are considering zero baryon density. In Fig.\ref{fg.distT}, we have shown $f_g$ and $f_q$ as functions of momenta at a temperature $T=T_c$. The corresponding distributions for ideal BE and FD statistics with the corresponding temperature dependent masses is also shown for comparison. Obviously, all the statistics melt to MB statistics at high momentum. For lower momenta, the quasi-gluons are found to be more suppressed from the ideal BE statistics than the quasi-quarks from the ideal FD statistics. This is the result of the lower value of $\langle \Phi_A \rangle$ than that of $\langle \Phi \rangle$ as seen in Fig.\ref{fig:placeholder}. This would have important consequences for the transport coefficients as discussed below. For completeness, we also show the comparative plots of the various distribution functions as functions of varying $T/T_c$ for a given momentum $p=0.9$ GeV in Fig.~\ref{fg.dist}.\\
%%%%%%%%%%
%%%%%%%%%%%%%%%%%%%%%%%%%%%%%%%%%%%%%%%%%%%%%%%%%%%%%%%%%%

\subsection{\label{ssec:trans}Transport coefficients}

We now present the shear ($\eta$) and bulk ($\zeta$) viscosities employing the analytic expressions derived within the relaxation time approximation~\cite{Chakraborty:2010fr, Albright:2015fpa}. For the quasi-quarks we have,
%%%%%%%%%%%%%%%%%%%%%%%%%%%%%%%%%%%%%%%%%%%%%%%%%%%%%%%%%%
\begin{align}
    &\eta_q=\frac{2 d_q\beta}{15}\int \frac{d^3p}
    {(2\pi)^3}\tau \left(\frac{|\vec{p}|^2}
    {\varepsilon_q}
    \right)^2 f_q(1-f_q),
    \label{eq_shearquark}
    \\
   &\zeta_q=2 d_q \beta \int \frac{d^3p\tau}{(2\pi)^3} 
   \left(\frac{\left(\frac{1}{3}-c_s^2\right)|\vec{p}|^2-
   c_s^2\frac{d}{d\beta^2}(\beta^2m^{*2})}{\varepsilon_q}
   \right)^2\nonumber\\
   &\quad \quad \quad\quad\quad\times  f_q(1-f_q),
   \label{eq_zetaquark}
\end{align}
%%%%%%%%%%%%%%%%%%%%%%%%%%%%%%%%%%%%%%%%%%%%%%%%%%%%%%%%%
and for quasi-gluons we have,
%%%%%%%%%%%%%%%%%%%%%%%%%%%%%%%%%%%%%%%%%%%%%%%%%%%%%%%%%%
\begin{align}
     &\eta_g=\frac{d_g\beta}{15}\int \frac{d^3p}
     {(2\pi)^3}
     \tau \left(\frac{|\vec{p}|^2}{\varepsilon_g}
     \right)^2 
     f_g(1+f_g),
    \label{eq_sheargluon}
   \\
   &\zeta_g= d_g \beta \int \frac{d^3p \tau}{(2
\pi)^3} 
   \left(\frac{\left(\frac{1}{3}-c_s^2\right)|\vec{p}|^2-
   c_s^2\frac{d}{d\beta^2}(\beta^2m_g^2)}
{\varepsilon_g}
   \right)^2\nonumber\\
   &\quad \quad\quad\quad\quad\times f_g(1+f_g)~,
   \label{eq_zetagluon}
    \end{align} 
%%%%%%%%%%%%%%%%%%%%%%%%%%%%%%%%%%%%%%%%%%%%%%%%%%%%%%%%%% 
where $\tau$ is the relaxation time and $\beta = 1/T$. 
%%%%%%%%%%%%%%%%%%%%%%%%%%%%%%%%%%%%%%%%%%%%%%%%%%%%%%%%%%
\begin{figure}[!htb]
{\includegraphics[width=\linewidth]
{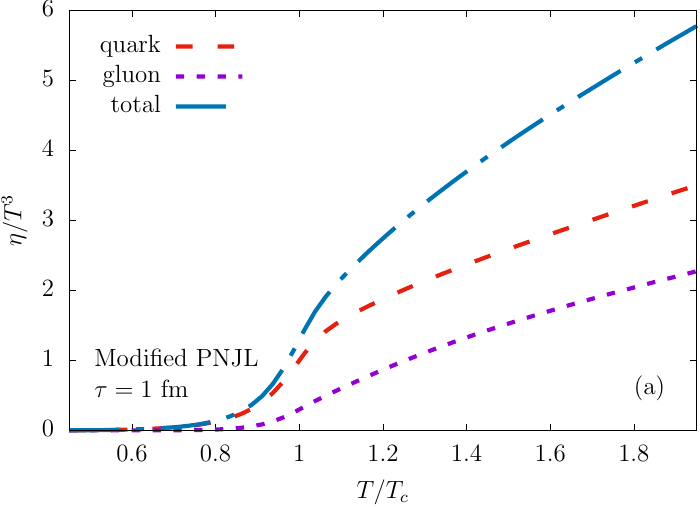}}
\\
{\includegraphics[width=\linewidth]
{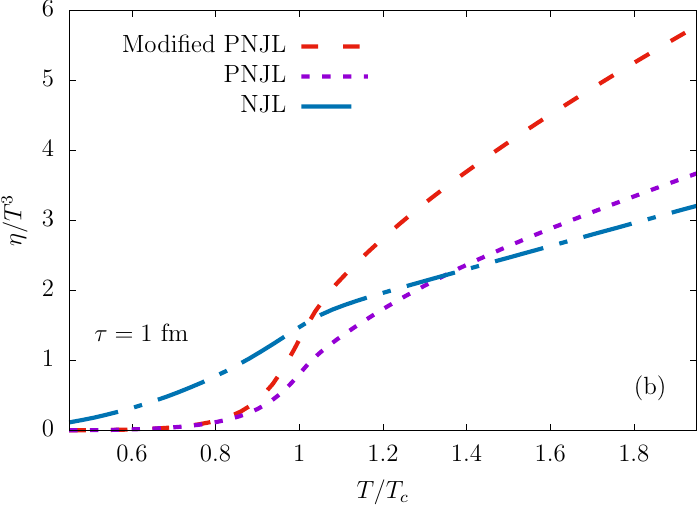}}
\caption{Temperature variation of shear viscosity. (a) Quark and gluonic contribution to the total $\eta$ for the modified PNJL model, (b) Comparison of the results from the current model to the results obtained in PNJL and NJL model.}
\label{fg.trans}
\end{figure} 
%%%%%%%%%%%%%%%%%%%%%%%%
\begin{figure}[!htb]
{\includegraphics[width=\linewidth]
{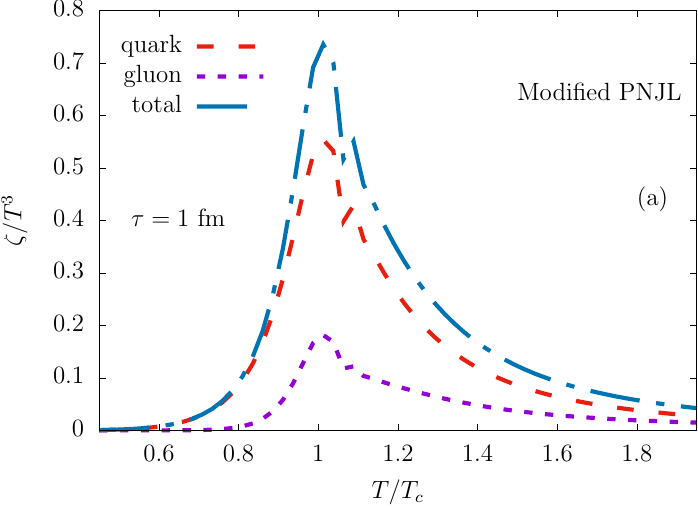}}\\
{\includegraphics[width=\linewidth]
{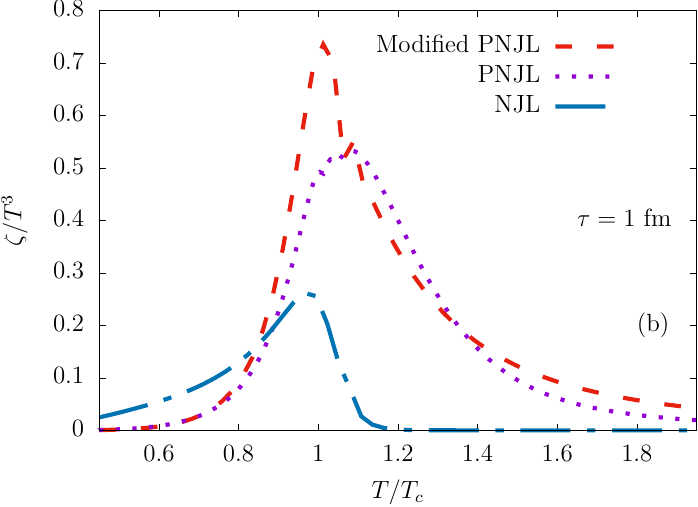}}
\caption{Temperature variation of bulk viscosity. (a) Quark and gluonic contribution to the total $\zeta$ for the modified PNJL model, (b) Comparison of the results from the current model to the results obtained in PNJL and NJL model.}
\label{fg.transbulk}
\end{figure}
%%%%%%%%%%%%%%%%%%%%%%%%%%%%%%%%%%%%%%
%%%%%%%%%%%%%%%%%%%%%%%%%%%%%%%%%%%%%%%%%%%%%%%%%%%%%%%%%%%%%
We begin our analysis with a temperature independent relaxation time $\tau= 1$ fm. This emphasizes the effect of phase space distribution and the thermodynamic quantities on the thermal behavior of the transport coefficients which can otherwise be  overshadowed by a temperature dependent relaxation time as we will see later. In Fig.\ref{fg.trans} we present the thermal behavior of $\eta/T^3$ as obtained in our model framework and compared with the results obtained within the usual NJL and PNJL description. To obtain the results for the standard PNJL model we have considered the quark distribution function as,
%%%%%%%%%%%%%%%%%%%%%%%%%%%%%%%%%%%%%%%%%%%%%%%%%%%
\begin{align}
  f_q(p_1,T)=\frac{e^{-3\frac{ \varepsilon_q(p_1)}{T}}+
(\Phi+2\bar{\Phi} e^{-\frac{ \varepsilon_q(p_1)}{T}}) 
e^{-
\frac{ \varepsilon_q(p_1)}{T}}}{1+e^{-3
\frac{ \varepsilon_q(p_1)}{T}}
+N_c(\Phi+\bar{\Phi}e^{-
\frac{ \varepsilon_q(p_1)}{T}})e^{-
\frac{\varepsilon_q(p_1)}{T}}} 
\end{align}
%%%%%%%%%%%%%%%%%%%%%%%%%%%%%%%%%%%%%%%%%%%%%%%%%%%%%%%%%
where $\Phi$ and $\bar{\Phi}$ and $\sigma$ are taken to be the corresponding mean field values evaluated from Eq.~\eqref{saddle}. For the NJL model the quark distribution function is the usual Fermi-
Dirac distribution function with the temperature dependent constituent quark mass defined as, $m^*=m_0+\sigma$. Here $\sigma$ is considered to be the mean field value, $\sigma_{\mathrm{mf}}$, obtained from,
%%%%%%%%%%%%%%%%%%%%%%%%%%%%%%%%%%%%%%%%%%%%%%%%%%%%%%%%%
\begin{align}
    \frac{\partial \Omega_q^{\Phi,\Bar{\Phi}
    \rightarrow 1}}{\partial \sigma}\bigg|_{\sigma=
    \sigma_{\mathrm{mf}}}=0,
\end{align}
%%%%%%%%%%%%%%%%%%%%%%%%%%%%%%%%%%%%%%%%%%%%%%%%%%%%%%%%%
where $\Omega_q$ is defined in Eq.~\eqref{eqomegaPNJL}. The model parameters used for the usual PNJL and NJL model are taken from table. \ref{table_u} and table.\ref{table}. Note that any studies of transport coefficients in standard PNJL model~\cite{FUKUSHIMA2004277, Megias:2004hj, PhysRevD.73.014019, Ghosh:2006qh, Megias:2006bn, Mukherjee:2006hq, Ghosh:2007wy} and NJL model~\cite{Nambu:1961tp, Nambu:1961fr, Hatsuda:1994pi, Vogl:1991qt, Klevansky:1992qe, Buballa:2003qv} do not take into account the phase space contribution from the quasi-gluon.

In Fig.\ref{fg.trans}(a), we present the scaled shear viscosity as a function of $T/T_c$. The separate quasiparticle contribution from quark and gluon sectors to the shear viscosity are also shown. Figure.\ref{fg.trans}(b), shows a comparative plot of the quasiparticle PNJL model estimation for scaled $\eta$ with the standard PNJL and NJL models. For lower temperatures both the standard and quasiparticle PNJL models produce $\eta/T^3$ which is highly suppressed compared to the NJL model due to the confining effect of background Polyakov loop. At higher temperatures the values in the the standard PNJL and NJL models approach each other, while that in the quasiparticle PNJL model is much higher than both of the other model estimates. A quick look at Fig.\ref{fg.trans}(a) clearly shows that this higher value is mainly because of the quasi-gluons present in the modified PNJL model, which establishes the main objective of this exercise.\\
In this context let us now discuss our findings for the bulk viscosity. As can be seen from Eq.\eqref{eq_zetaquark} and Eq.\eqref{eq_zetagluon}, for the bulk viscosity, apart from the distribution functions we have important contribution from the pre-factor that depends on the speed of sound and thereby on the departure from conformality. The NJL model has a harder equation of state than the two PNJL models and therefore is expected to have a lower bulk viscosity. On the other hand the gluon sector is primarily responsible for the departure from conformality. So the question is what would be its implication in the quasiparticle PNJL model. 
Figure.\ref{fg.transbulk} shows $\zeta/T^3$ as function of $T/T_c$. In Fig.\ref{fg.transbulk}(a) we show our model results where the quark contribution dominates. 
The peak around the transition temperature is related to the deviation from the conformal limit. As can be seen from Eq.\eqref{eq_zetaquark} and Eq.\eqref{eq_zetagluon} $\zeta$ increases as speed of sound deviates from the conformal limit $1/3$ and the quasiparticle mass shows a sharp transition around $T_c$. On the other hand at the low temperature system moves towards the non relativistic limit, and the suppression of degrees  of freedom drags the coefficient down. Similar to $\eta/T^3$ in fig.\ref{fg.transbulk}(b) we see due to the confinement effect $\zeta/T^3$ is smaller for the Polyakov models compared to the NJL results.
%%%%%%%%%%%%%%%%%%%%%%%%
\begin{figure}
{\includegraphics[width=\linewidth]{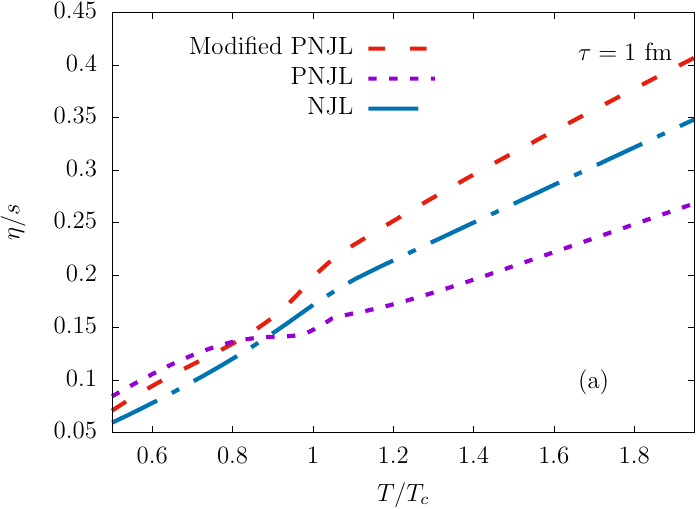}} \\{\includegraphics[width=\linewidth]{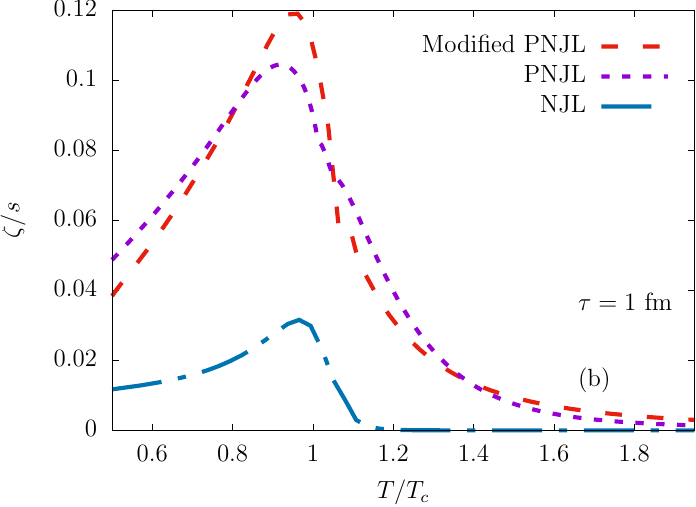}} 
	\caption{Temperature variation of $\eta/s$ (a) and $\zeta/s$ (b) calculated within the current model and compared to the results obtained in PNJL and NJL model for constant $\tau$.}
	\label{fg.etabyScons}
\end{figure} 
%%%%%%%%%%%%%%%%%%%%%%%%%%%%%%%%%%%
\paragraph*{Estimation of $\eta/s$ and $\zeta/s$:} The dimensionless shear viscosity to entropy density ratio can be used to compare both relativistic and nonrelativistic fluids, and expected to show 
an abrupt change near the phase transition region~\cite{Schaefer:2014awa}. The reason for this is different interaction mechanism for momentum transport in different phases of the fluids~\cite{Kapusta:2008vb}. As can be seen From Fig.\ref{fg.etabyScons}, with a constant relaxation time, both $\eta/s$ and $\zeta/s$ show qualitatively similar behavior as Fig. \ref{fg.trans}(b) and Fig.\ref{fg.transbulk}(b) respectively. The $\eta/s$ ratio simply increases monotonically with temperature, while the $\zeta/s$ ratio rises with temperature up to $T_c$ and then decreases. These behavior come only from the phase space factor. Therefore it is now pertinent to ask what would be the behavior if one takes into account the temperature variation of the relaxation time. While computation of the relaxation time for the three models is beyond the scope of the present study a heuristic analysis may be done considering the facts that the relaxation time is expected to be high in the sparse hadronic phase and typically decrease with increasing temperature. Beyond the critical temperature the relaxation time is expected to stabilize or rise slowly with the interaction strength of the deconfined quarks and gluons falling off. Keeping these limiting cases in mind, we considered a generic form for the temperature dependent relaxation time for two flavor QGP medium obtained within the NJL model. The energy averaged temperature dependent relaxation time, $\tau=g(T)$, for quark-quark and quark-antiquark scattering shows a minima close to chiral crossover temperature ($T_c$)
$\sim 1.0487T_c$.~\cite{Deb:2016myz}.  However, in our current model this relaxation time is not directly connected to such specific interactions; rather we considered it as a generic relaxation time for all the constituents that incorporates all possible interactions within the medium. The temperature dependent relaxation time $\tau(T)$ considered here is the one shown in Fig.6a of Ref.~\cite{Deb:2016myz} for zero baryon chemical potential.
%%%%%%%%%%%%%%%%%%%%%%%%%%%%%%%%%%%%%%%%%%%%%%%%%%%%%%%%%%
\begin{figure}[!htb]
{\includegraphics[width=\linewidth]{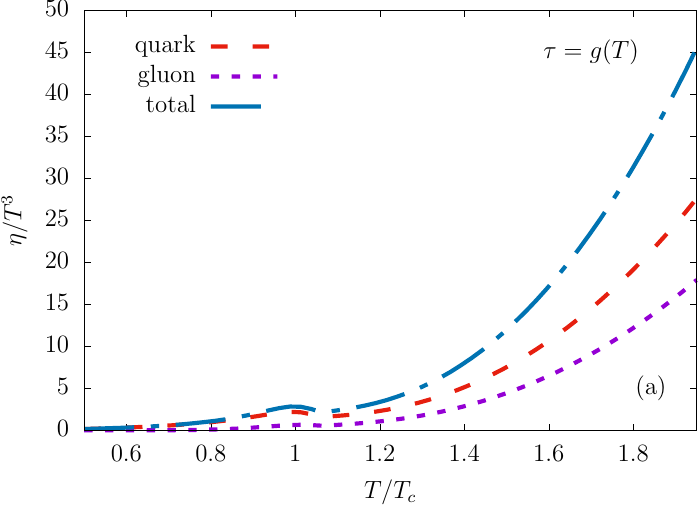}} \\
{\includegraphics[width=\linewidth]{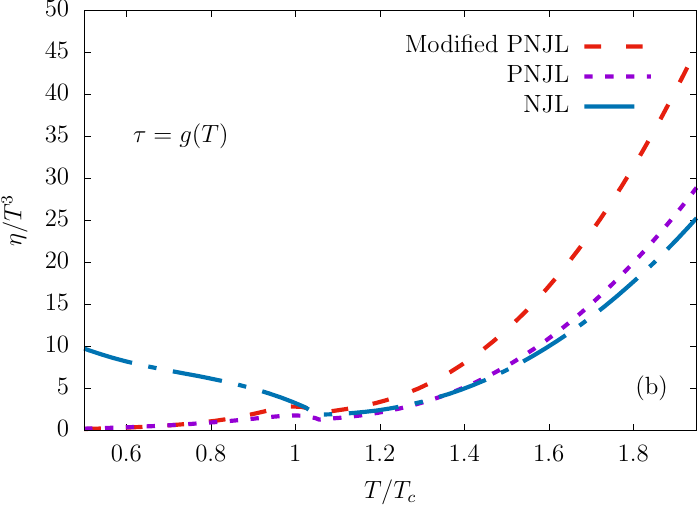}} 
	\caption{Temperature variation of shear viscosity with temperature dependent relaxation time. (a) Quark and gluonic contribution to the total $\eta$ for the modified PNJL model. (b) Comparison of the results from the current model to the results obtained in PNJL and NJL model.}
	\label{fg.etabyT3param}
\end{figure} 
%%%%%%%%%%%%%%%%%%%%%%
\begin{figure}[!htb]
	{\includegraphics[width=\linewidth]
	{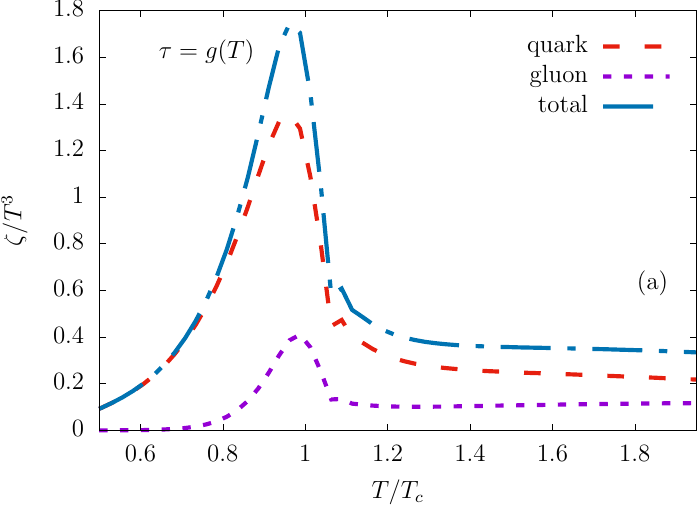}} \\
    {\includegraphics[width=\linewidth]
	{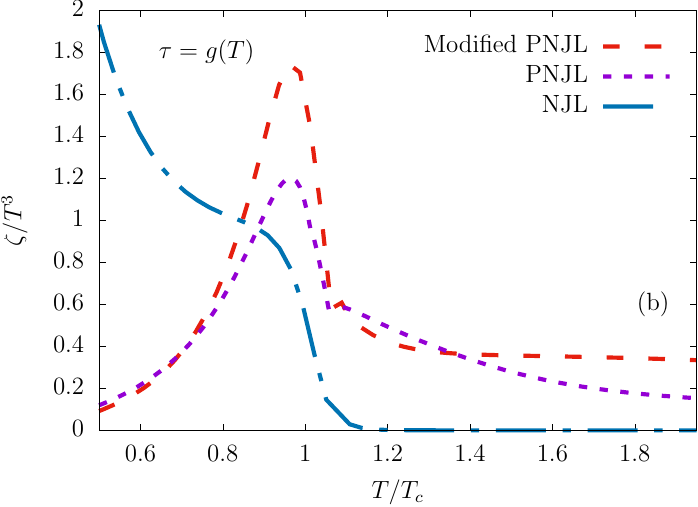}} 
	\caption{Temperature variation of bulk viscosity with temperature dependent relaxation time. (a) Quark and gluonic contribution to the total $\zeta$ for the modified PNJL model. (b) Comparison of the results from the current model to the results obtained in PNJL and NJL model.}
	\label{fg.zetabyT3param}
\end{figure} 

%%%%%%%%%%%%%%%%%%%%%%%%%
\begin{figure}[!htb]
	{\includegraphics[width=\linewidth]
	{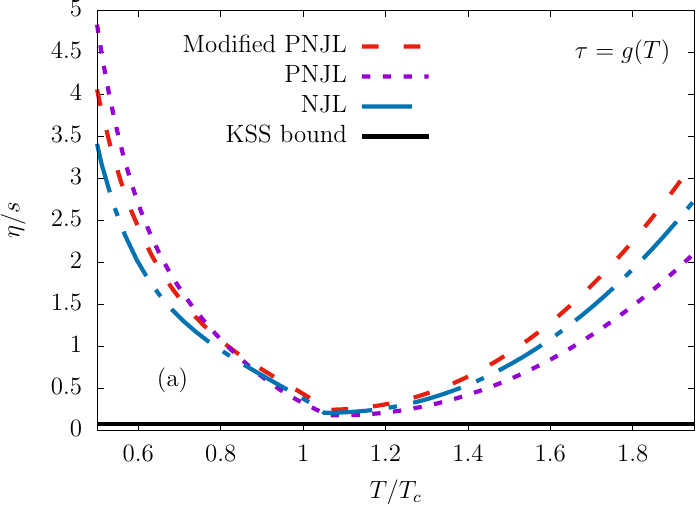}} \\
    {\includegraphics[width=\linewidth]
	{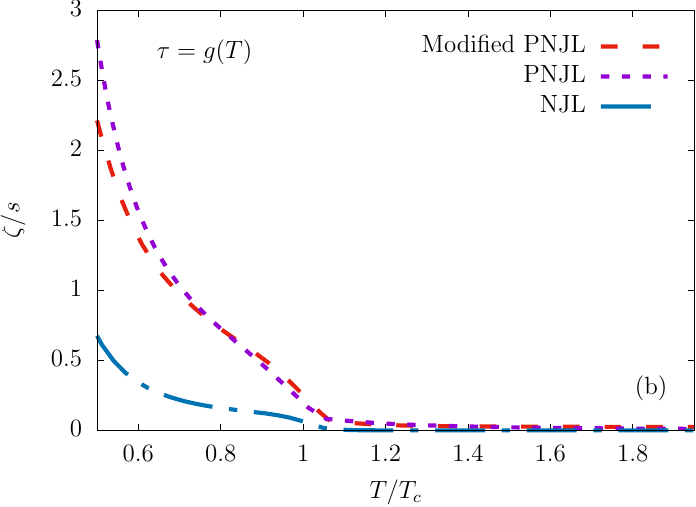}} 
	\caption{Temperature variation of $\eta/s$ (a) and $\zeta/s$ (b) calculated within the current model and compared to the results obtained in PNJL and NJL model.}
	\label{fg.etabyS}
\end{figure} 
%%%%%%%%%%%%%%%%%%%%%%
%%%%%%%%%%%%%%%%%%%%

Considering this form of the temperature dependent relaxation time we first show $\eta/T^3$ and $\zeta/T^3$ in Fig. \ref{fg.etabyT3param} and Fig.\ref{fg.zetabyT3param} respectively. One can see that for NJL model the temperature dependent $\tau$ dominates over the 
suppression in the degrees of freedom due to the fact that the constituent quarks are not sufficiently suppressed at low temperatures. On the other hand for the Polyakov loop extended models the confining effects ensures that at the low temperatures both the transport coefficients decrease with decreasing temperature. In the high temperature region for $\eta/T^3$ the interaction effects enhance the increasing nature of the transport coefficient. However $\zeta/T^3$ maintains a peak structure projecting the dominant effect from deviation of conformality. Note that in both these ratios, as also in the case of constant relaxation times the contribution of the gluon quasiparticles are significant. At the same time there contribution is still lower than that of the quark quasiparticles due to the difference in the phase space contributions as discussed earlier.

Finally in fig.\ref{fg.etabyS} we present the $\eta/s$ and $\zeta/s$ as functions of $T/T_c$. We see the typical shear viscosity to entropy density ratio for all three models. Since the confinement effect in the degrees of freedom gets canceled by the entropy density, the interaction effects dominate, leading to increasing $\eta/s$ with decreasing temperature. The minima in the $\eta/s$ sit near but not exactly at the critical temperature which is a direct consequence of the parametric form of $\tau$ considered in this work. Similar behavior was also observed for NJL model in Ref.\cite{Deb:2016myz}.
Similarly the peak structure in the $\zeta/T^3$ ratio is completely overshadowed in the $\zeta/s$ ratio by the fast decrease of entropy with decreasing temperature. 
%%%%%%%%%%%%%%%%%%%%%%%%%%%%%%%%%%%%%%%%%%%%%%%%%%%%%%%%%%
%----------------------------------------------------------------------
\section{\label{sec:discussion}Discussion} 
%----------------------------------------------------------------------
In this article, we have introduced a quasi-particle effective model framework that consistently describes QCD thermodynamics for a two flavor system. We have coupled a quasi-gluon model with background Polyakov loop to the chiral effective model setup. Through the quasi-particle description, this model incorporates dynamical properties for both quark and gluon degrees of freedom where confinement properties are realized through background Polyakov loop.\\ 
The scope of the usual Polyakov extended chiral effective models is limited for studies where the distribution function for gluonic degrees of freedom becomes essential. 
As a more generalised approach a gluon quasiparticle model in the background Polyakov field can be derived in the SU(3) pure Yang-Mills theory within the background field method. Coupling that description to the chiral effective model provides a uniform QCD quasiparticle framework. However, the model was reported to have physical inconsistencies at temperatures below $T_c$. In our previous work we have addressed this issue within the pure gauge set up.  Instead of saddle point approximation  we obtained physically consistent results by incorporating an alternative integral  approach. In the present work, we applied the integral approach for dealing with the partition function of the quasi-particle PNJL model for the 
Polyakov fields while retaining the usual 
saddle point approach for the chiral field $\sigma$. In general the framework presented here opens up the potential to explore various physics problems where contributions from both quark and gluon quasi-particles are essential. 
%-----------------------------------------------------------------------
\begin{acknowledgments}
RR would like to thank the Department of Science and Technology (DST). MGM acknowledges the financial support from Department of Atomic Energy (DAE), Govt. of India through Raja Ramanna Chair Scheme. PS is funded by   the EU’s NextGenerationEU instrument through the National Recovery and Resilience Plan of Romania - Pillar III-C9-I8, managed by the Ministry of Research, Innovation and Digitization, within the project entitled ``Facets of Rotating Quark-Gluon Plasma'' (FORQ), contract no.~760079/23.05.2023 code CF 103/15.11.2022.  C.A.I. and P.S. would like to thank Kazuyuki Kanaya for providing the data of the paper~\cite{CP-PACS:2001hxw}. C.A.I. would also like to acknowledge the financial support by the Chinese Academy of Sciences President's International Fellowship Initiative (PIFI) under Grant No. 2020PM0064. A major part of the work was carried out while he was a PIFI fellow.
\end{acknowledgments}
%-----------------------------------------------------------------------

%\clearpage 

\bibliography{ref1}

\end{document}